\documentclass[pdflatex, sn-nature]{sn-jnl2}

\usepackage{textgreek}

\jyear{2021}%

\theoremstyle{thmstyleone}%

\theoremstyle{thmstyletwo}%

\theoremstyle{thmstylethree}%

\newcommand\jname{JADES-GS+53.18343-27.79097}
\newcommand\jnamespace{JADES-GS+53.18343-27.79097 }

\newcommand{\apjl}{Astrophys. J. Lett.}   
\newcommand{\apjs}{Astrophys. J. Suppl. Ser.}   
\newcommand{\ao}{Appl. Opt.}   
\newcommand{\aap}{Astron. Astrophys.}   

\begin{document}

\title[Inside-out growth in the early Universe]{A core in a star-forming disc as evidence of inside-out growth in the early Universe}








\author*[1,2]{\fnm{William} \sur{M.~Baker}}\email{wb308@cam.ac.uk}
\author*[1,2]{\fnm{Sandro} \sur{Tacchella}}\email{st578@cam.ac.uk}
\author[3]{\fnm{Benjamin} \sur{D.~Johnson}}
\author[4]{\fnm{Erica} \sur{Nelson}}
\author[]{\fnm{Katherine} \sur{A.~Suess$^{\textit{\normalfont{5,6}}}$}} 
\author[1,2]{\fnm{Francesco} \sur{D'Eugenio}}
\author[1,2,7]{\fnm{Mirko} \sur{Curti}}
\author[8]{\fnm{Anna} \sur{de Graaff}}
\author[9]{\fnm{Zhiyuan} \sur{Ji}}
\author[1,2,10]{\fnm{Roberto} \sur{Maiolino}}
\author[5]{\fnm{Brant} \sur{Robertson}} 
\author[1,2]{\fnm{Jan} \sur{Scholtz}}

\author[9]{\fnm{Stacey} \sur{Alberts}}
\author[11]{\fnm{Santiago} \sur{Arribas}} 
\author[12,13]{\fnm{Kristan} \sur{Boyett}}
\author[14]{\fnm{Andrew} \sur{J.~Bunker}}
\author[15]{\fnm{Stefano} \sur{Carniani}}
\author[16]{\fnm{Stephane} \sur{Charlot}}
\author[9]{\fnm{Zuyi} \sur{Chen}}
\author[14]{\fnm{Jacopo} \sur{Chevallard}}
\author[17]{\fnm{Emma} \sur{Curtis-Lake}}
\author[1,2]{\fnm{A. Lola} \sur{Danhaive}}
\author[9]{\fnm{Christa} \sur{DeCoursey}}
\author[9]{\fnm{Eiichi} \sur{Egami}}
\author[3]{\fnm{Daniel} \sur{J.\ Eisenstein}}
\author[18]{\fnm{Ryan} \sur{Endsley}} 
\author[19]{\fnm{Ryan} \sur{Hausen}}
\author[9]{\fnm{Jakob} \sur{M.~Helton}}
\author[20]{\fnm{Nimisha} \sur{Kumari}}
\author[1,2]{\fnm{Tobias} \sur{J.~Looser}}
\author[21]{\fnm{Michael} \sur{V.~Maseda}}
\author[1,2]{\fnm{D\'avid} \sur{Pusk\'as}}
\author[9]{\fnm{Marcia} \sur{Rieke}}
\author[1,2]{\fnm{Lester} \sur{Sandles}}
\author[9]{\fnm{Fengwu} \sur{Sun}}
\author[1,2]{\fnm{Hannah} \sur{\"Ubler}}
\author[22]{\fnm{Christina} \sur{C.~Williams}}
\author[9]{\fnm{Christopher} \sur{N.~A.~Willmer}}
\author[1,2]{\fnm{Joris} \sur{Witstok}}

\affil[1]{Kavli Institute for Cosmology, University of Cambridge, Madingley Road, Cambridge, CB3 OHA, UK}

\affil[2]{Cavendish Laboratory - Astrophysics Group, University of Cambridge, 19 JJ Thomson Avenue, Cambridge, CB3 OHE, UK}

\affil[3]{Center for Astrophysics $\|$ Harvard \& Smithsonian, 60 Garden St., Cambridge, MA 02138, USA}

\affil[4]{Department for Astrophysical and Planetary Science, University of Colorado, Boulder, CO 80309, USA}

\affil[5]{Department of Astronomy and Astrophysics, University of California, Santa Cruz, 1156 High Street, Santa Cruz, CA 96054, USA}

\affil[6]{Kavli Institute for Particle Astrophysics and Cosmology and Department of Physics, Stanford University, Stanford, CA 94305, USA}

\affil[7]{European Southern Observatory, Karl-Schwarzschild-Strasse 2, D-85748 Garching bei Muenchen, Germany}

\affil[8]{Max-Planck-Institut f\"ur Astronomie, K\"onigstuhl 17, D-69117, Heidelberg, Germany}

\affil[9]{Steward Observatory University of Arizona 933 N. Cherry Avenue ,Tucson, AZ 85721, USA}

\affil[10]{Department of Physics and Astronomy, University College London, Gower Street, London WC1E 6BT, UK}

\affil[11]{Centro de Astrobiolog\'ia (CAB), CSIC–INTA, Cra. de Ajalvir Km.~4, 28850- Torrej\'on de Ardoz, Madrid, Spain}

\affil[12]{School of Physics, University of Melbourne, Parkville 3010, VIC, Australia}

\affil[13]{ARC Centre of Excellence for All Sky Astrophysics in 3 Dimensions (ASTRO 3D), Australia}

\affil[14]{Department of Physics, University of Oxford, Denys Wilkinson Building, Keble Road, Oxford OX1 3RH, UK}

\affil[15]{Scuola Normale Superiore, Piazza dei Cavalieri 7, I-56126 Pisa, Italy}

\affil[16]{Sorbonne Universit\'e, CNRS, UMR 7095, Institut d'Astrophysique de Paris, 98 bis bd Arago, 75014 Paris, France}

\affil[17]{Centre for Astrophysics Research, Department of Physics, Astronomy and Mathematics, University of Hertfordshire, Hatfield AL10 9AB, UK}

\affil[18]{Department of Astronomy, University of Texas, Austin, TX 78712, USA}

\affil[19]{Department of Physics and Astronomy, The Johns Hopkins University,  3400 N. Charles St., Baltimore, MD 21218, USA}

\affil[20]{AURA for European Space Agency, Space Telescope Science Institute, 3700 San Martin Drive, Baltimore, MD 21218, USA}

\affil[21]{Department of Astronomy, University of Wisconsin-Madison, 475 N. Charter St., Madison, WI 53706, USA}

\affil[22]{NSF's National Optical-Infrared Astronomy Research Laboratory, 950 North Cherry Avenue, Tucson, AZ 85719, USA}


\abstract{The physical processes that establish the morphological evolution and the structural diversity of galaxies are key unknowns in extragalactic astrophysics. Here we report the finding of the morphologically-mature galaxy JADES-GS+53.18343-27.79097, which existed within the first 700 million years of the Universe's history. This star-forming galaxy with a stellar mass of 400 million solar masses consists of three components, a highly-compact core with a half-light radius of less than 100 pc, an actively star-forming disc with a radius of about 400 pc, and a star-forming clump, which all show distinctive star-formation histories. The central stellar mass density of this galaxy is within a factor of two of the most massive present-day ellipticals, while being globally 1000 times less massive. The radial profile of the specific star-formation rate is rising toward the outskirts. This evidence suggests the first detection of inside-out growth of a galaxy as a proto-bulge and a star-forming disc in the Epoch of Reionization.}





\maketitle

\section*{}\label{sec1}

In the hierarchical $\Lambda$CDM cosmological model, galaxies sustain their star formation for extended periods of time in a quasi-steady state of gas inflow, gas outflow, and gas consumption \cite{Bouche2010, Dekel2013}. To first order, the gas that cools at later cosmic epochs possesses higher angular momentum; therefore, it settles in a more extended star-forming disc, implying that galaxies grow from the inside out \cite{Fall1980, Mo1998}. However, the actual formation of galaxies in the cosmological context is more complex since a wide range of processes regulate star formation and the orbital distribution of stars, ranging from stellar feedback (from supernovae and stellar winds), black hole feedback, and cosmic rays, to galaxy-galaxy interactions and mergers \cite{Van-den-Bosch2001, Agertz2011, Marinacci2014, Tacchella2016, Girichidis2018}. Therefore, the morphological structure and spatially resolved growth rates of galaxies are a sensitive -- but also complicated -- probe of galaxy formation physics \cite{Ubler2014,Sales2010, Dutton2009, Dubois2016, Cochrane2023, Scannapieco2012, Naab2017}. 

Galaxies in the local Universe display a range of morphologies, from younger disc-dominated spiral galaxies to older bulge-dominated ellipticals \cite{Kormendy2004, Simard2011}, and are typically classified by the Hubble sequence \cite{Hubble1926,Kormendy1996,galaxy_zoo_10.1093/mnras/stt1458}. The growth of local star-forming galaxies has been observed on spatially resolved scales, confirming that generally galaxies grow inside out \cite{Munoz-Mateous2007,Pezzulli2015, Frankel2019}. However, there is a diverse range of specific star-formation rate (sSFR) profiles in the local universe with some galaxies undergoing inside-out growth, while others grow outside-in \cite{Nelson2016, Belfiore2018, Ellison2018}, likely corresponding to different growth phases.
Most of the mass of local galaxies is found to have formed during the redshift range $1\leq z \leq3$, around the period of ``cosmic noon'', the peak of the cosmic star formation rate density in the Universe \cite{Madau2014}. Observations at these redshifts have revealed many galaxies with massive bulges and rotating discs \cite{Tacchella2015, Lang2014, Nelson2014, Nelson2016}. However, in order to probe the buildup of these $1\leq z \leq3$ bulges, we need to investigate even earlier cosmic times, characterising galaxies during the Epoch of Reionization ($z\gtrsim6$, \cite{Robertson2022}). Observationally, little is known about how quickly these early galaxies grow, in particular on spatially resolved scales.
The theoretical expectation is that the galaxy merger rate increases towards higher redshifts \cite{Rodriguez-gomez2015, Fakhouri2010, Snyder2017}, which could lead to more pronounced central starbursts \cite{Hernquist1989, Mihos1996}. 
Recent numerical models indicate that early (at $z>3$), low-mass ($10^{7-9}~M_{\odot}$) galaxies can undergo rapid size fluctuations and compaction driven by the competition between feedback-driven gas outflows and cold inflows \cite{El-Badry2016, Shen2024}, leading to a flat or even inverted size-mass relation with more massive galaxies being smaller ($<10^9~M_{\odot}$). Direct observations will address questions about how galaxies grow their stellar mass and size in the early Universe, thereby providing insights into the physics that regulates star formation.

JWST opens a new window to study the formation of the Hubble sequence and bulge-disc formation in the early Universe \cite{Karpaltepe2023, Suess2022, Ferreira2023}. As part of the JWST Advanced Deep Extragalactic Survey (JADES, \cite{Eisenstein2023, Robertson2023, Curtis-Lake2023}), we report here the discovery of a core-disc galaxy with an off-centre clump (\jname) at a spectroscopic redshift of 7.430 (see Methods), when the Universe was only 700 million years old. \jnamespace appears to be growing inside-out, having built up a massive compact core at its centre before forming a surrounding star-forming disc. This is the first time we are able to characterise a core-disc system during the Epoch of Reionization and find the signature of early bulge formation.

\begin{figure}
    \centering
    \includegraphics[width=0.45\columnwidth]{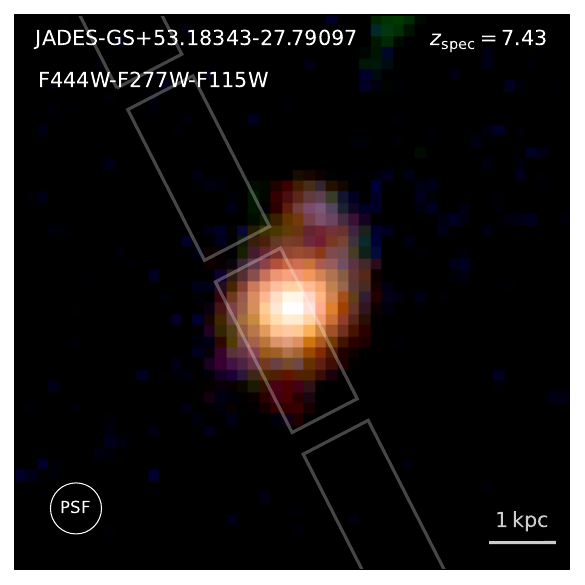}
    \includegraphics[width=0.53\columnwidth]{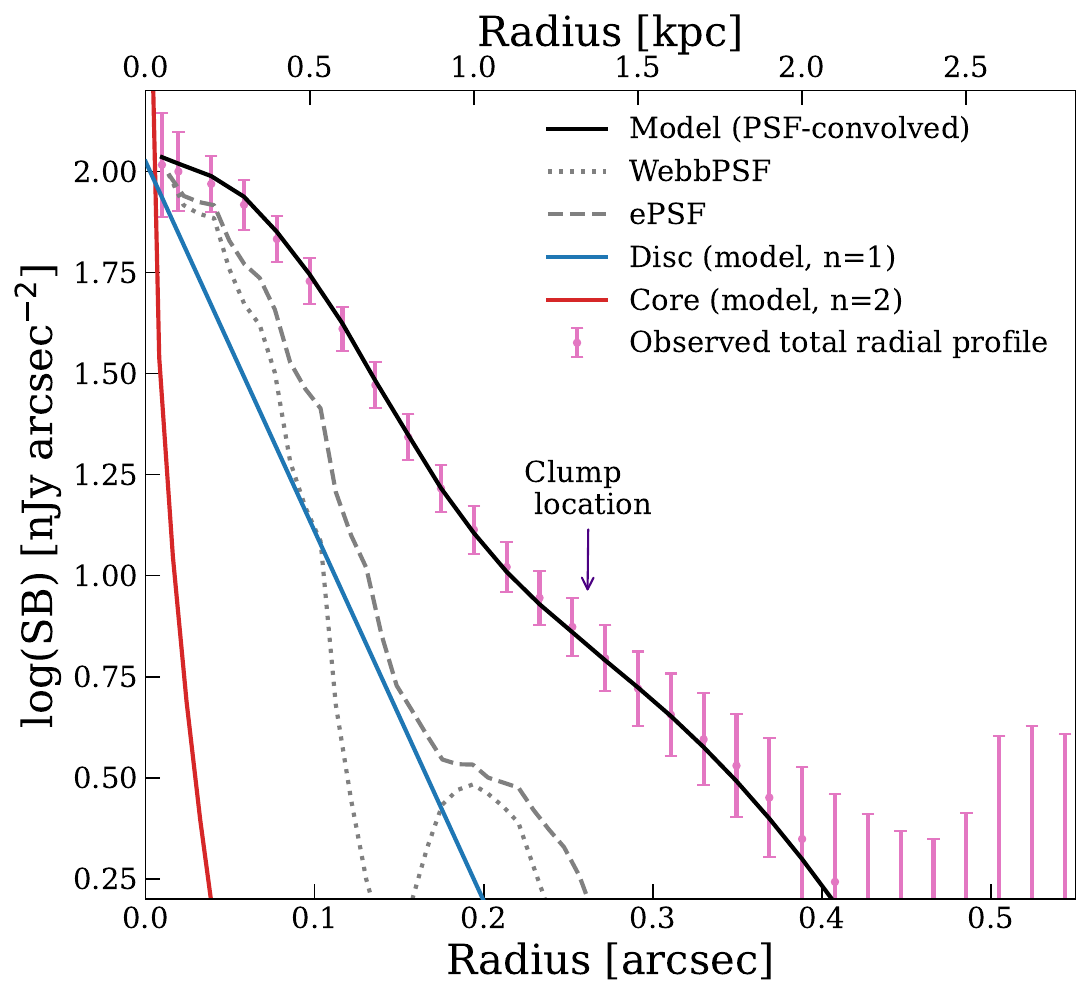}
    \includegraphics[width=\columnwidth]{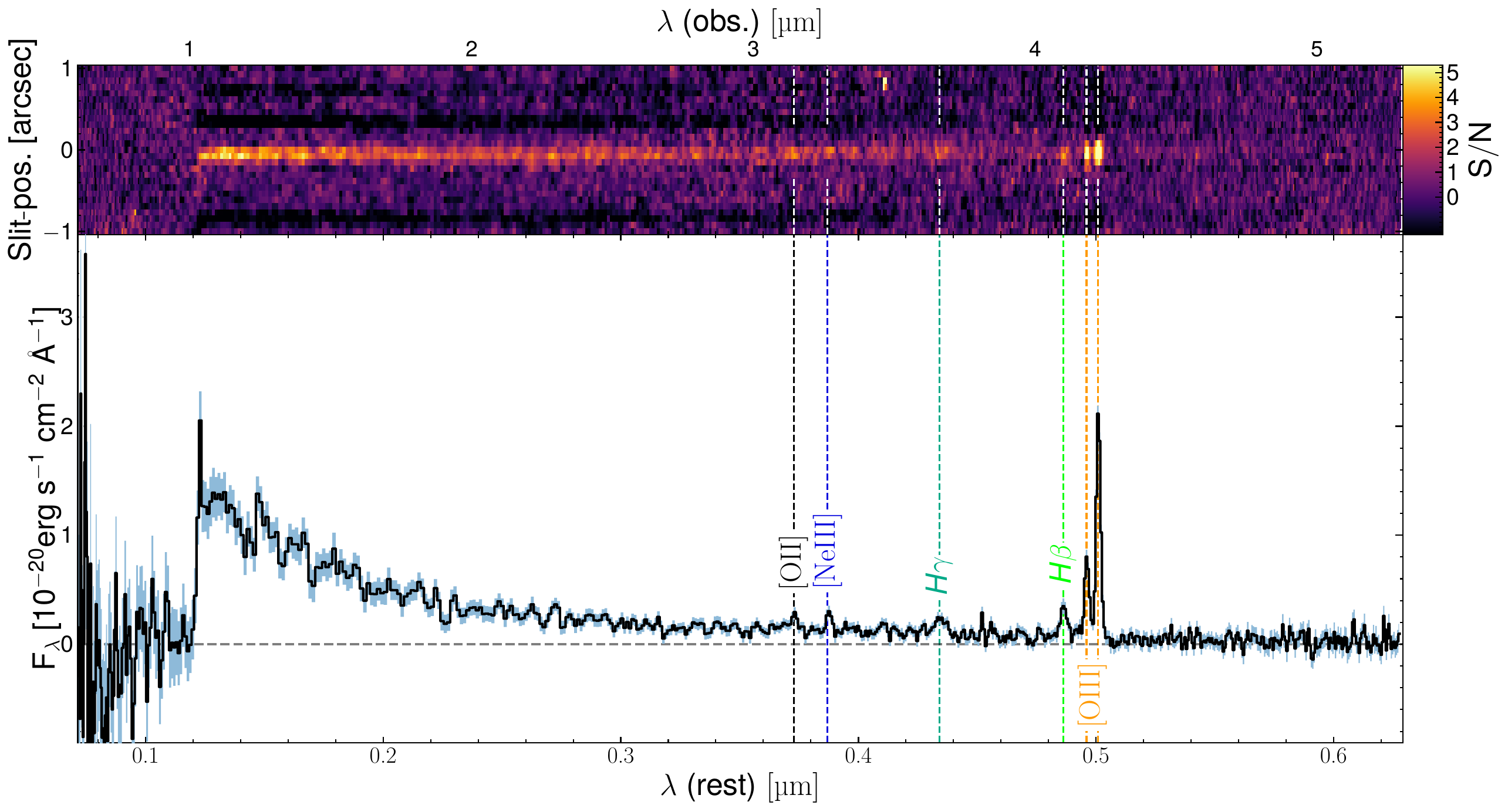}
    \caption{\textbf{Image of the galaxy, radial profile of the light distribution and the 1D and 2D spectra.} Upper left panel: the (F444W-F277W-F115W) colour composite image of the galaxy, where the central core and disc are prominent. A 1-kpc size bar (corresponding to 0.19 arcsec), the F444W PSF and the position of the NIRSpec slit are overplotted. 
    Upper right panel: the total radial (azimuthally averaged) surface brightness profile of the core, disc and clump in the F356W band, where the pink points are the observational data and the black line is the PSF-convolved best-fit three-component model (consisting of the core, disc and clump). 
    The errorbars correspond to the error propagated through from the error maps (i.e. the standard deviation).
    The intrinsic best-fit S\'ersic profiles of the core and disc components are shown as red and blue lines, respectively. The off-centred clump at a distance of 1.4 kpc is indicated with a purple arrow. The grey dashed and dotted lines correspond to the empirical PSF (ePSF) and the WebbPSF, respectively, in the F356W band. 
    Lower panel: the 2D and 1D NIRSpec R100 prism spectra, with the position of notable detected emission lines overplotted, which indicates that this galaxy is dominated by stellar emission. The errors on the spectrum correspond to the standard deviation of the mean signal. }
    \label{fig:rgb}
\end{figure}


\subsection*{Core-disc-clump decomposition}

\begin{figure*}
    \centering
    \includegraphics[width=0.49\columnwidth]{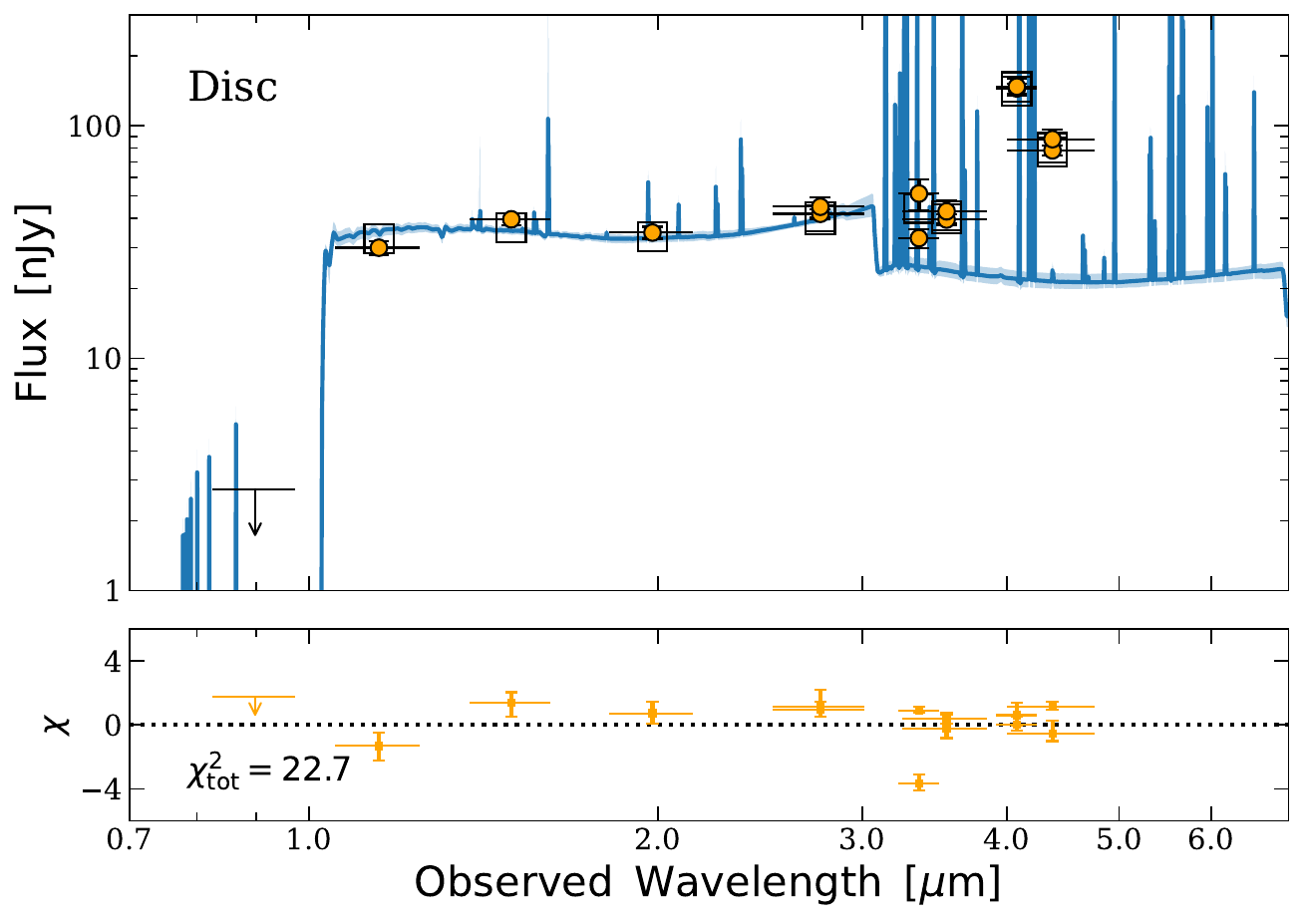}
    \includegraphics[width=0.49\columnwidth]{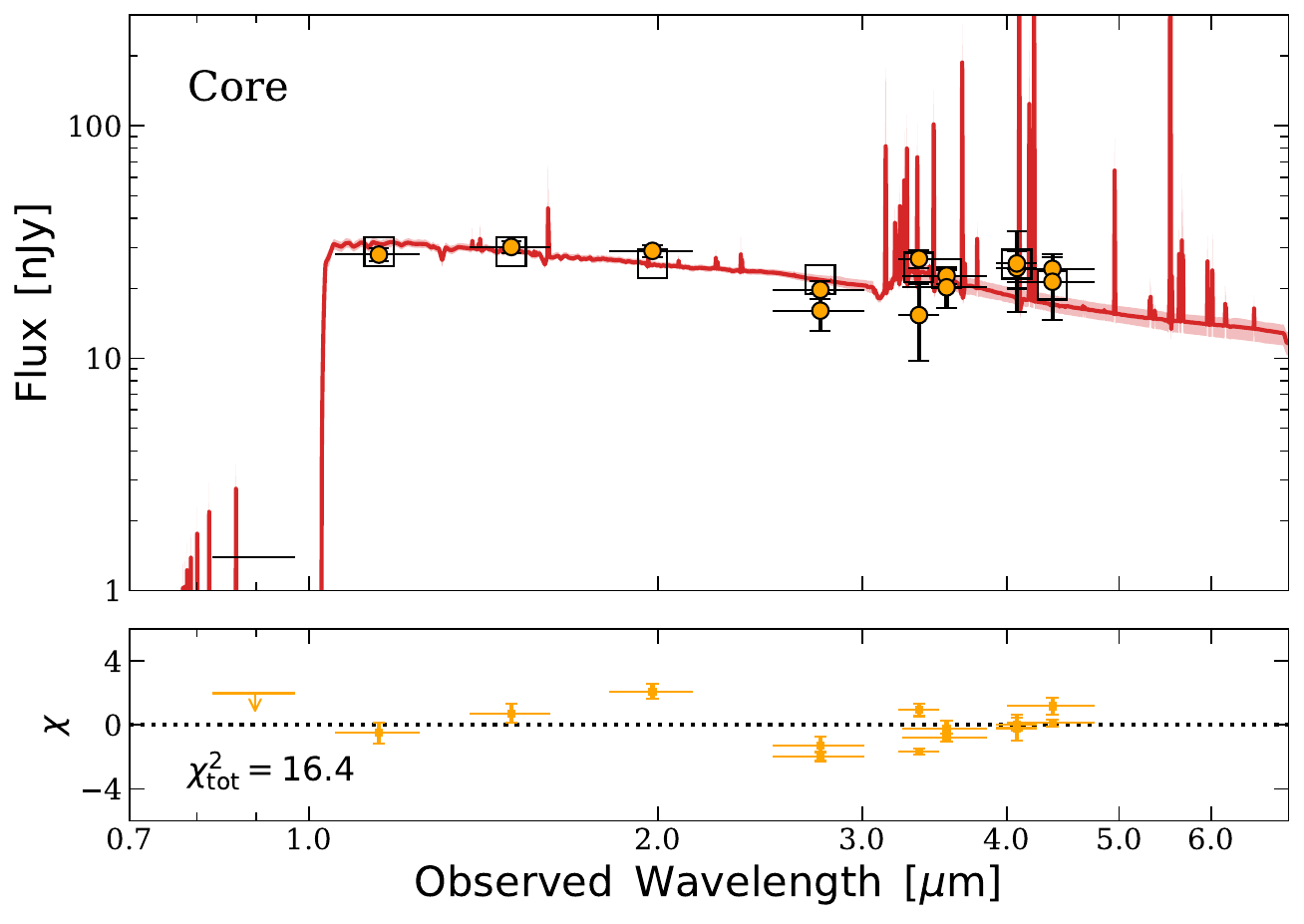}
    \includegraphics[width=0.49\columnwidth]{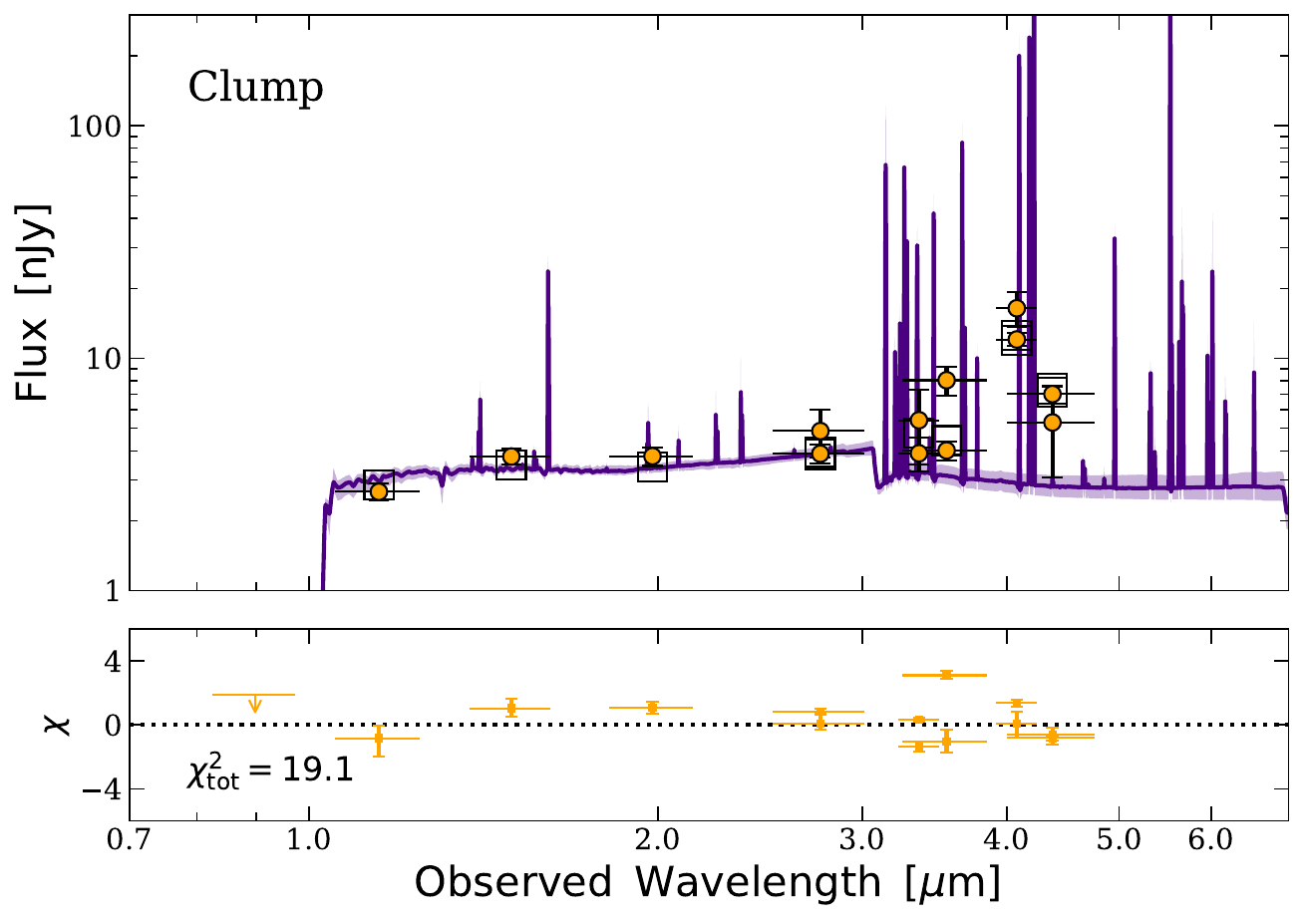}
    \includegraphics[width=0.49\columnwidth]{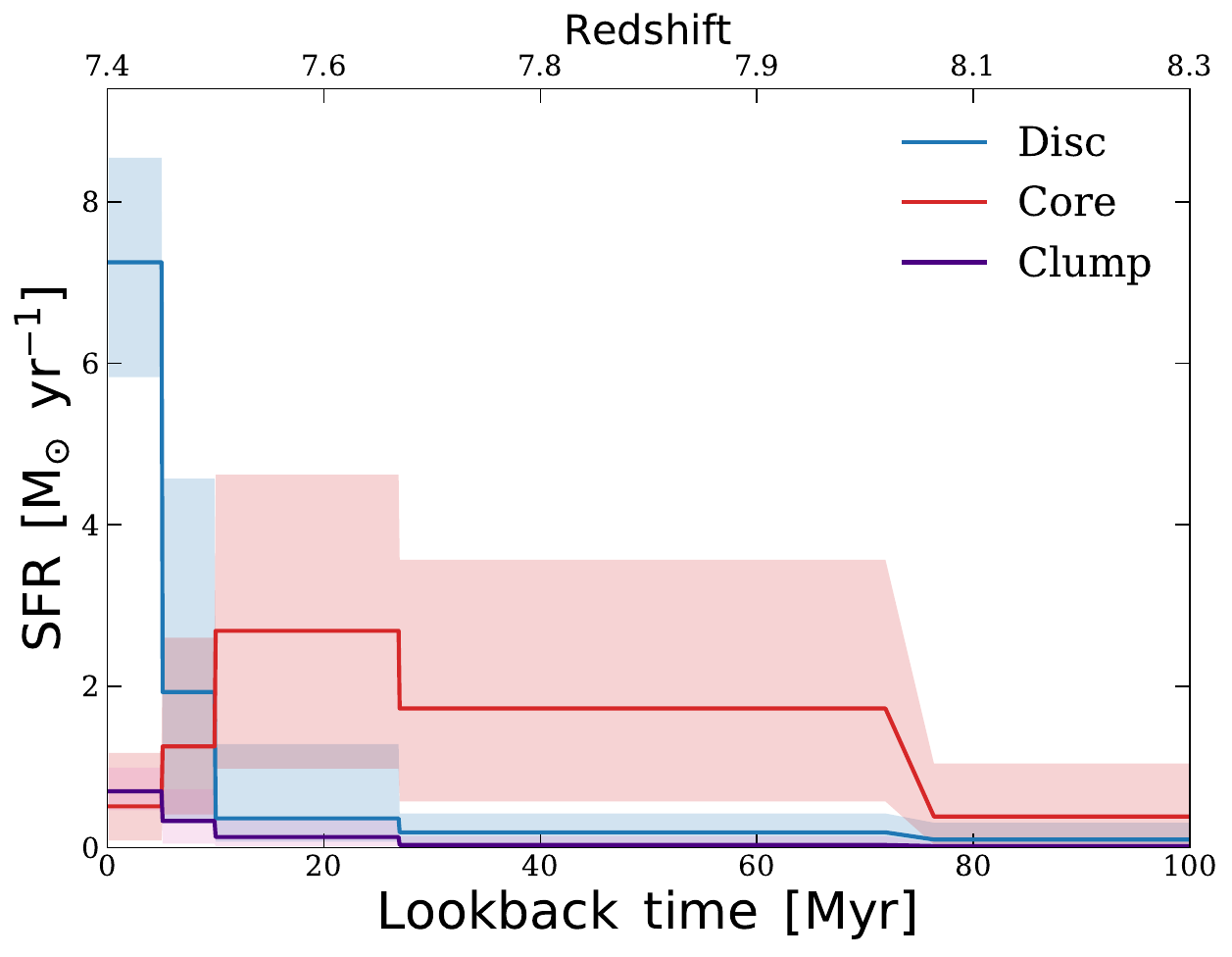}
    \caption{\textbf{Spectral energy distributions for the three components and their star-formation histories.} Spectral energy distribution (SED) fits for the disc (upper left), core (upper right), and clump (lower left) components. The yellow points show the photometry inferred from the ForcePho modelling, while the 2$\sigma$ upper limits are indicated as downward pointing arrows. The errors on the photometry are the 1$\sigma$ uncertainties (which are derived via standard error propagation through the reduction pipeline). The errors on the x-axis correspond to the widths of the filters.
    The open squares mark the photometry of the best-fit SED model. The solid lines and the shaded regions show the median and the 16th-84th percentile of the SED posterior from the Prospector modelling.
    Lower right: star-formation rate (SFR) against look-back time (from the observed redshift) for the disc (blue line), core (red line) and clump (purple line). The shading corresponds to the 16th and 84th percentiles of the star-formation history posterior.
    We find that the disc has been undergoing a burst of star formation for the last 5 Myr, while the core is entering a lower-star-formation phase. The clump has a lower SFR compared to the other two components and is over an order of magnitude less massive than the core.}
    \label{fig:seds}
\end{figure*}

We use NIRCam \cite{Rieke2023} imaging in nine filters (F090W, F115W, F150W, F200W, F277W, F335M, F356W, F410M and F444W) and NIRSpec/MSA \cite{Ferruit2022, Jakobson2022, Boker2023} spectroscopy from JADES in the GOODS-S region \cite{Eisenstein_goodss_proposal_jadesprop}. This gives us extended coverage of the rest-frame ultra-violet and optical wavelengths (including the Balmer break), which constrains the stellar populations on spatially resolved scales. Furthermore, the medium band F410M probes the strength of the emission lines H$\beta$+[OIII] on spatially resolved scales.

The upper left panel of Fig. \ref{fig:rgb} shows a colour-composite red-green-blue (RGB, corresponding to F444W-F277W-F115W) image of \jname. The image shows a compact central component (core) surrounded by an extended, disc-like component. \jnamespace has a strong colour gradient between the central region and the outskirts (Methods Figure \ref{fig:image_color}), implying an excess of H$\beta$+[OIII] in the outskirts as probed by F410M. In order to quantify the compactness and colour gradient, we employ the tool ForcePho (B. Johnson, in prep.) to perform a detailed morphological and photometric analysis of \jname, forward modelling all individual exposures across all bands simultaneously and accounting for the point-spread functions (PSFs; see Methods, Section \ref{sec:forcepho}). We explore a range of different models and find that a three-component model, consisting of a disc S\'ersic profile (fixing the S\'ersic index to $n=1$ by limiting the bounds to $\pm 0.01$), a compact component ($n=2-5$) consistent with a bulge or pseudobulge, and an off-centre clump (modelled as a quasi-point source), reproduces the data best (with the smallest residuals and $\chi^2$). Although this model is not unique, it is able to capture the complex morphology of this galaxy in all of the filters.
We obtain a central core with a half-light radius of 16 mas (roughly 80 pc), while the disc has a half-light radius of 80 mas (about 400 pc; see Table \ref{tab:table}), overall confirming the compact nature of the source. Although we refer to the extended component as a ``disc'' from a purely morphological perspective, the ionised gas (as traced by F410M) of \jnamespace is actually consistent with a rotationally supported component with $v(r_{\rm eff})/\sigma_0\approx1.3$ \cite{DeGraaff2023}.

Fig. \ref{fig:rgb}, upper right panel, shows the surface brightness profile of the galaxy in the F356W band, the best-fit model convolved with the PSF (black), the unconvolved S\'ersic model components (core in red and disc in blue), and the WebbPSF and empirical PSF (ePSF) of the mosaic (grey dotted and dashed line). This shows that the PSF-convolved three-component model reproduces the observed surface brightness profile (see also Methods, Section \ref{sec:1v2} for a detailed plot of the model and residual). We also fit a single-component model with and without the clump, plus just a core + disc fit, all of which we test against our fiducial three-component model. We find that all of these model variants fail to account for either the additional flux in the centre and/or the flux of the clump, resulting in higher $\chi^2$ statistic values (0.45, 0.25, and 0.18 vs 0.14 for the fiducial three-component fit; see Methods, Section \ref{sec:1v2}). We also calculate a reduced $\chi^2$ which favours the fiducial three-component model. To check against PSF-approximation issues with ForcePho, we re-simulate the core, disc, and clump fits convolved with the WebbPSF, and then refit them. Using the WebbPSF model, we find that the results are consistent with the original fit within the uncertainties, confirming that the ForcePho PSF approximations are appropriate (see Methods, Section \ref{sec:psf_approx}).

Fig. \ref{fig:rgb}, bottom panel, shows the 2D and 1D NIRSpec R100 prism spectra of \jname, including the positions of notable detected emission lines. This spectrum probes both the core and the disc, as indicated by the slit position in the upper left panel of Fig. \ref{fig:rgb}. Using these data, we estimate a spectroscopic redshift of $z=7.430$, consistent with the photometric redshift. The measured emission line fluxes from the NIRSpec prism spectrum indicate that this galaxy is consistent with stellar emission, and the high-resolution grating spectrum shows that all emission lines (in particular H$\beta$) are narrow, implying that this galaxy does not appear to host a dominant Active Galactic Nucleus (AGN; see Methods, Section \ref{sec:SF_AGN}), but we note that it is difficult to fully rule out an AGN.

\subsection*{Stellar population properties}

\begin{table}
    \centering
    \resizebox{\columnwidth}{!}{%
    \begin{tabular}{llllllllll}
\hline
       & $\log$(M$_{\star}/M_{\odot}$)   & t$_{\rm half}$      & SFR$_{\rm 10Myr}$            & SFR$_{\rm 100Myr}$           & $\log(Z/Z_\odot)$          & $\rm A_{\rm V}$     & n         & $r_e$           \\
        &  & [Myr]  & $\rm [M_{\odot} yr^{-1}]$ & $\rm [M_{\odot} yr^{-1}]$ & &   & & [mas] \\
\hline
 Disc  & ${7.98}_{-0.19}^{+0.25}$          & ${19}_{-15}^{+108}$ & ${4.8}_{-0.6}^{+0.7}$                 & ${0.7}_{-0.2}^{+0.3}$                  & ${-1.5}_{-0.3}^{+0.5}$ & ${0.08}_{-0.03}^{+0.05}$ & ${1.0^*}$ & ${80}_{-16}^{+16}$       \\
 Core  & ${8.38}_{-0.16}^{+0.18}$          & ${68}_{-31}^{+78}$  & ${1.0}_{-0.7}^{+0.8}$                 & ${1.7}_{-0.7}^{+0.8}$                  & ${-1.7}_{-0.2}^{+0.4}$ & ${0.02}_{-0.02}^{+0.07}$ & ${2.03}_{-0.3}^{+0.3}$ & ${16}_{-3}^{+3}$       \\
 Clump & ${7.22}_{-0.22}^{+0.42}$          & ${44}_{-38}^{+146}$ & ${0.5}_{-0.1}^{+0.2}$                 & ${0.1}_{-0.0}^{+0.1}$                  & ${-1.3}_{-0.4}^{+0.7}$ & ${0.13}_{-0.06}^{+0.09}$ & ${1.0^*}$ & ${5}_{-0}^{+0}$        \\
\hline
\end{tabular}
}
    \caption{Table showing (from left to right), stellar mass, half-time (the half-mass assembly time), star formation rate (SFR) averaged over 10 Myr, SFR averaged over 100Myr, stellar metallicity, extinction in the V-band $\rm A_{\rm V}$, S\'ersic index $n$, and half light radius r$_e$, for the core, disc and clump components from our fiducial fit using ForcePho. The S\'{e}rsic index of the disc and clump are fixed to $n=1$ (indicated by the star).}
    \label{tab:table}
\end{table}

We show in Fig.~\ref{fig:seds} the ForcePho-based spectral energy distributions (SEDs) of the core, disc, and clump components.
To explore the stellar population properties of the three components, we fit the individual SEDs using the Bayesian SED-fitting tool Prospector \cite{Prospector2021}. We input the flux values and errors obtained for each band from ForcePho, and independently fit the SEDs with a flexible star-formation history (SFH) with the standard continuity prior \cite{Leja2019}, a variable dust attenuation law with a free dust attenuation law index and normalisation, and a nebular emission model. We also test using a bursty continuity prior for the SFH, finding that we obtain stellar masses consistent with the standard continuity prior (see Methods, Section \ref{sec:SEDFitting}). In Fig. \ref{fig:seds} we plot the best fit for the SEDs of the three components, indicating that the SEDs are well reproduced by stellar emission in conjunction with dust attenuation and nebular emission.

Table \ref{tab:table} gives the key stellar population properties of the core, disc, and clump components. The core is the most massive of the components with log($\rm M_*/M_\odot)=8.4_{-0.2}^{+0.2}$, while the disc has log($\rm M_*/M_\odot)=8.0_{-0.2}^{+0.3}$, despite the core having a radius ($\sim80$ pc) less than a quarter of that of the disc ($\sim400$ pc). This suggests that the core is dense (stellar mass surface density of $\Sigma_{\rm eff}=\frac{M_*}{2\pi\,r^2}\approx 6\times10^{9}~M_{\odot}~\mathrm{kpc}^{-2}$). The bottom right panel of Figure \ref{fig:seds} shows the star-formation rate (SFR) as a function of lookback time for the core (red), the disc (blue) and the clump (purple). These SFHs show the full posterior distributions, i.e., are taking into account the degeneracies with other parameters, such as the dust attenuation law. The core, disc and clump have varying SFHs, with the core undergoing an earlier period of star formation with a recent decline, while the disc is currently undergoing a burst of star formation. Consistently, the stellar age (lookback time when half of the stellar mass formed) for the core is rather old with $t_{\rm half}=68^{+78}_{-31}$ Myr, while the disc is younger ($t_{\rm half}=19^{+108}_{-15}$ Myr). In total, we find that the combined core+disc galaxy has a stellar mass of log($M_*/M_\odot)=8.5^{+0.2}_{-0.2}$, and SFR$\rm_{\rm 10Myr}=5.8^{+1.5}_{-1.3} M_\odot/yr$, giving it a specific SFR (sSFR) of log(sSFR/yr)=$-7.7_{-0.1}^{+0.1}$ (typical for $7 \leq z \leq 8$ galaxies \cite{Endsley2023}). For comparison, from the NIRSpec spectrum we derive a dust-corrected H$\beta$ SFR of $\mathrm{SFR}_{\rm H\beta}=9_{-7}^{+30}~M_{\odot}/\mathrm{yr}$, which is consistent with our SED-derived SFR over 10 Myr. We find a similar agreement for dust attenuation (Balmer decrement) and gas-phase metallicity between the SED-derived and NIRSpec-derived values (see Methods, Section \ref{sec:em_lines}). Although the clump is not the focus of this study, we find it to be low in stellar mass (log($M_*/M_\odot)=7.2_{-0.2}^{+0.4}$) and highly star forming with a specific SFR (sSFR) of $\rm log(sSFR/yr^{-1})=-7.5^{+0.3}_{-0.4}$ averaged over 10 Myr. The clump's age appears to be relatively unconstrained ($t_{\rm half}=44^{+146}_{-38}$ Myr), meaning that it could be young and have been formed through a disc instability, or -- alternatively -- it could be older and be an accreted satellite galaxy.

In summary, we find that the core and disc are similarly bright in the rest-UV and that the disc is dominating the most recent star formation and is slightly dust-attenuated. 
This is surprising in the $z\approx2$ picture of a red dusty bulge embedded in a blue disc with lower dust attenuation, but consistent with recent JWST observations showing populations of red dusty discs missed in previous rest-optically selected samples \cite{Nelson2023}. The observational data that drive this result is the higher F410M excess in the disc (see Methods, Figure \ref{fig:image_color}), which implies strong nebular line contribution and is consistent with inside-out growth. This is a good showcase of the power of medium-band observations to constrain stellar populations \cite{Williams2023}.

\begin{figure}
    \centering
    \includegraphics[width=\columnwidth]{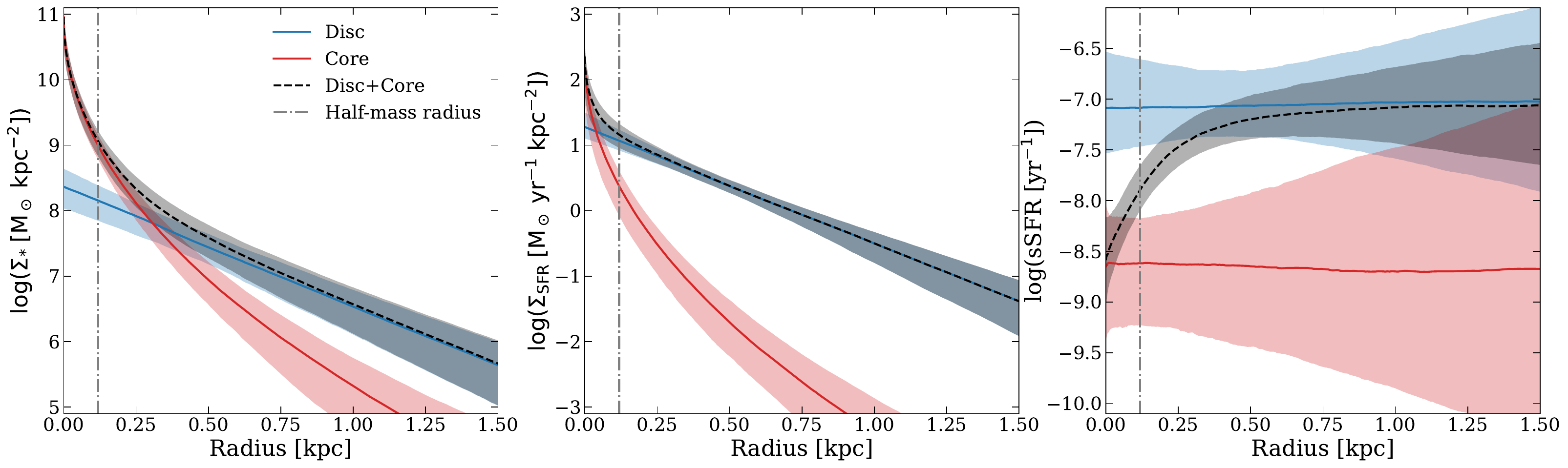}
    \caption{\textbf{Radial surface density profiles of stellar mass and star formation.} Radial profiles of stellar mass surface density ($\Sigma_{*}$; left panel), star-formation rate (SFR) surface density ($\Sigma_{\rm SFR}$; middle panel), and specific SFR (sSFR; right panel) for the core (solid red lines), disc (solid blue lines) and the combination of the two (dashed black lines). The shaded regions show the 16th and 84th percentile uncertainties from both the SED fitting and the structural measurements. The half-mass radius at the time of observation is overplotted as a vertical dashed-dotted line. This figure shows that the SFR surface density of the galaxy is, at almost all radii, completely dominated by the star-forming disc, while the stellar mass surface density in the central regions is dominated by the core. The sSFR increases with radius, implying that this galaxy grows from the inside out.}
    \label{fig:surface-density-radial-profiles}
\end{figure}

\subsection*{Radial profiles of stellar mass and star formation}

The compact size and the medium-band excess with increasing radius (probing the youngest stars via nebular emission) are indicative of a high central stellar mass density with a star-forming outskirt. We now use our multi-wavelength morphological decomposition and stellar population modelling to derive the intrinsic radial profiles of the stellar mass and SFR surface density. We use the unconvolved best-fit S\'ersic profiles, normalised to the best-fit stellar mass and SFR for each component, as given (for example) for the stellar mass surface density profile by $\rm \Sigma_{*}(r)=\frac{M_*\; I(r)}{I_{tot}}$, where I(r) is the intensity inferred from the S\'ersic profile at radius r and $\rm I_{tot}=\int 2\pi\, I(r)\,r\,dr$. The assumption is that each component has a negligible radial gradient in their stellar populations. However, since the normalisation of the individual components vary as a function of wavelength, we are able to account for stellar population gradients across the galaxy.

Fig. \ref{fig:surface-density-radial-profiles} shows the stellar mass surface density ($\Sigma_{*}$, left panel), the SFR surface density ($\Sigma_{\rm SFR}$, middle panel), and the sSFR (right panel) against the radius for the core and disc components and the combined profile. The $\Sigma_{\rm SFR}$ profile shows how the disc completely dominates the profile compared to the core, while the $\Sigma_{*}$ profile shows that the core's stellar mass surface density is prominent in the inner regions. We indicate in each diagram as a vertical dot-dashed grey line the half-mass radius ($R_{\star}=126^{+37}_{-26}$ pc), which is the radius of the galaxy at which half the (core+disc) stellar mass is contained. 

By construction, because they have the same S\'ersic profile (that is, shape) for the SFR and stellar mass radial profiles, the sSFR (= SFR/M$_*$) profiles are radially constant for the individual disc and core components, while their combined sSFR profile is rising with radius since $\mathrm{sSFR(r)} = \frac{\Sigma_{\rm SFR, core}+\Sigma_{\rm SFR, disc}}{\Sigma_{\rm *, core}+\Sigma_{\rm *, disc}}$. The sSFR profile shows where the galaxy grows, as the stellar mass doubling timescale is approximately equal to 1/sSFR. We see that the sSFR is steeply rising at the half-mass radius (about 1.5 dex within the central 1 kpc): the central 100 pc has a stellar mass doubling time of $\approx100$ Myr, while the outskirt has a mass doubling timescale of $\approx10$ Myr. This implies that this galaxy increases its half-mass radius with time and grows inside out.

\begin{figure*}
    \centering
    \includegraphics[width=0.47\columnwidth]{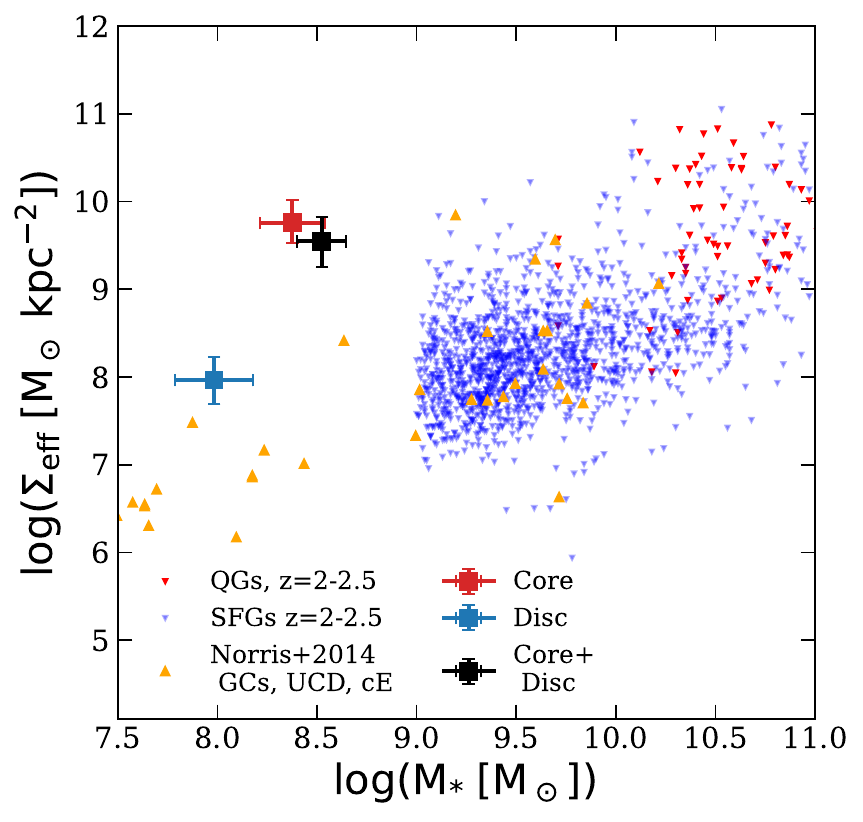}
    \includegraphics[width=0.49\columnwidth]{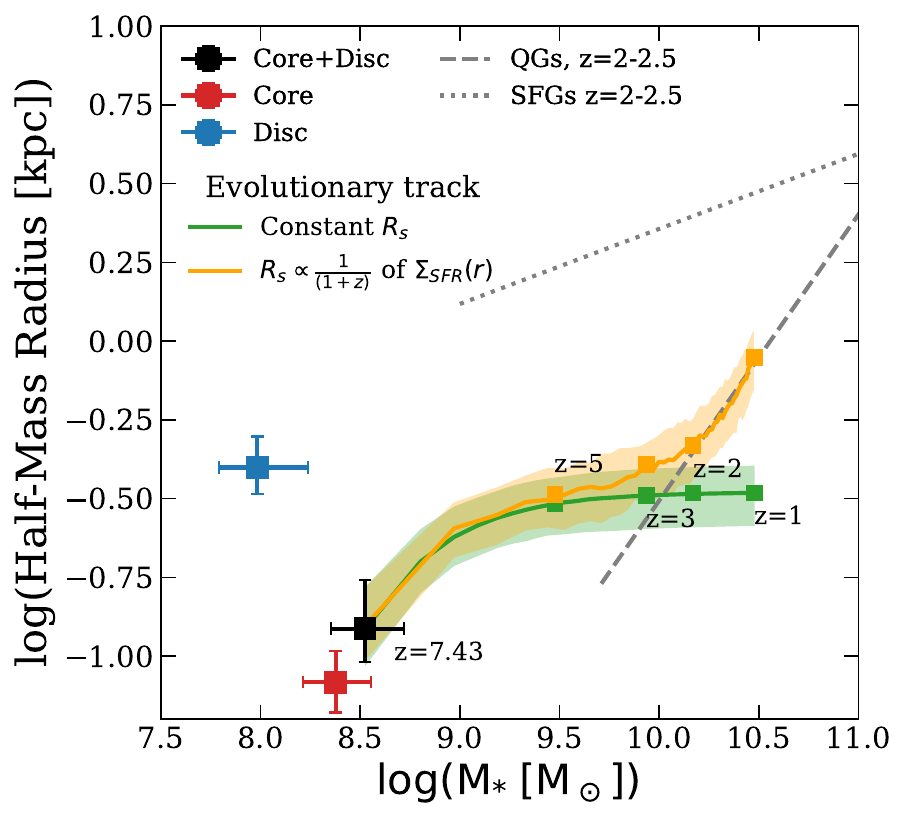}
    \caption{\textbf{Stellar mass density and half-mass size in the context of lower-redshift galaxies.} Left: effective stellar mass density ($\Sigma_{\rm eff}$) versus stellar mass for the core (red square), disc (blue square), and the combination of both components (black square). The data is presented as the median of the distribution with errors corresponding to the 16th and 84th percentiles.
    We compare this to measurements from $z\approx2$ \cite{Suess2019} and $z=0$ \cite{Norris2014}. Although our $z=7.43$ galaxy is low in stellar mass, $\Sigma_{\rm eff}$ lies in the upper envelope of $z\approx2$ star-forming galaxies (SFGs) and in the range of quiescent galaxies (QGs). Right: half-mass size versus stellar mass for the core (red square), disc (blue square), and the combination of both components (black square). The errors correspond to the 16th and 84th percentiles of the resulting distribution. The dotted and dashed lines show the size-mass relation at redshift $z\approx2$ \cite{Suess2019} for SFGs and QGs, respectively. The solid green line marks the predicted evolutionary track to $z=1$ assuming the inferred SFR profile (Fig.~\ref{fig:surface-density-radial-profiles}), while the orange line marks the evolutionary track for a model galaxy where the scale length of the SFR density of the core and disc scale as $1/(1+z)$. The half-mass radius at redshifts 5, 3, 2 and 1 is represented by the small squares, indicating that -- given the current growth rate -- both models predict that this galaxy will evolve onto the size-mass relation of quiescent galaxies at $z=2$.}
    \label{fig:structure}
\end{figure*}

\begin{figure}
    \centering
    \includegraphics[width=0.8\columnwidth]{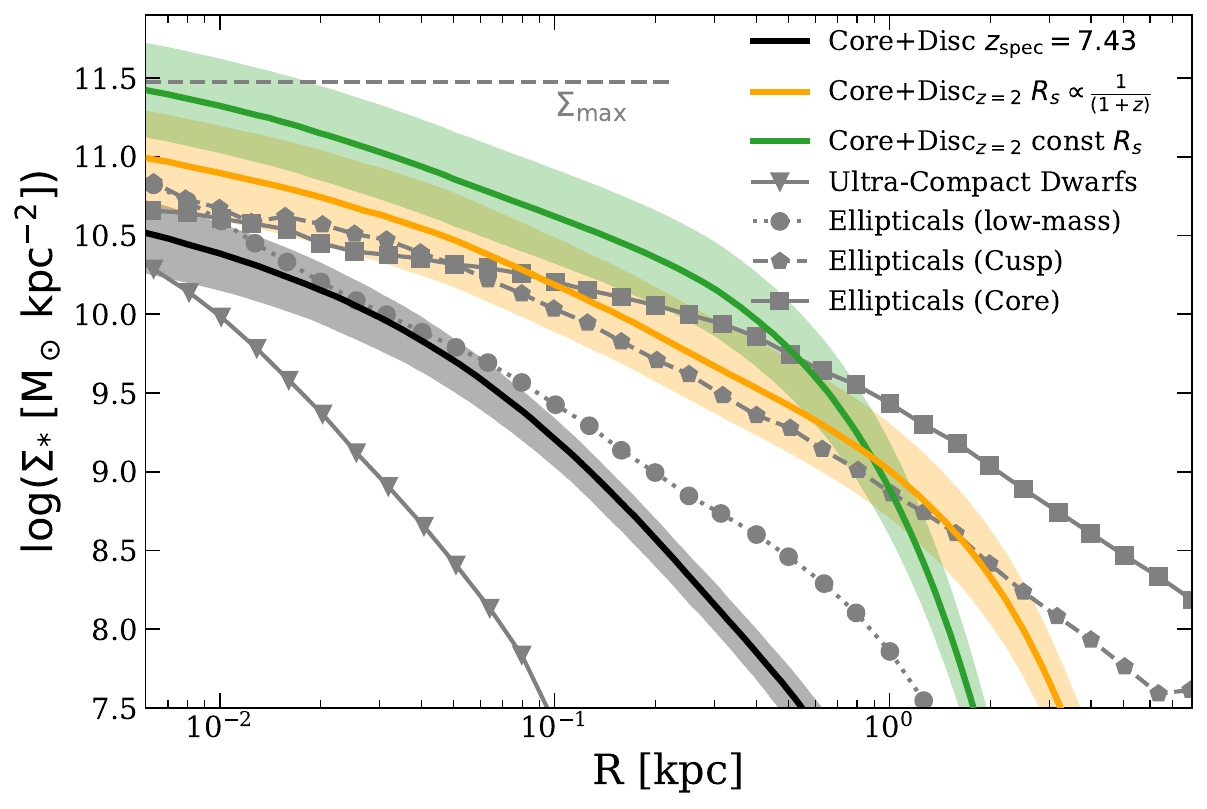}
    \caption{\textbf{Stellar mass surface density profile in comparison with local galaxies.} Radial profiles of stellar mass surface density ($\Sigma_{*}$) of JADES-GS+53.18343-27.79097 at $z_{\rm spec}=7.43$ (black line with shaded region corresponding to 16th and 84th percentiles) compared to average $z=0$ stellar mass density profiles of Ultra Compact Dwarfs (UCDs; grey line with triangles), low-mass ellipticals (grey dotted lines with points) and high-mass ellipticals (with core or cusp as a solid line with squares and a dashed line with squares respectively) \cite{Hopkins2010}. We find that this galaxy is extremely compact: the stellar mass density is consistent within 0.2-0.3 dex of today's massive ellipticals in the central region, while it is a factor of a 1000 lower in total stellar mass. Also overplotted are the extrapolated profiles at z=2, with the constant star-formation rate scale length ($R_{\rm s}$) model in green and the $R_s\propto \frac{1}{1+z}$ model in orange. We see that the redshift dependent scale model appears more reasonable compared with the redshift 0 ellipticals. }
    \label{fig:compact-density-radial-profiles}
\end{figure}

\subsection*{Discussion}

We now compare our $z=7.43$ galaxy with galaxies and stellar systems at lower redshift, which allows us to gain a complementary view on the spatially resolved growth from the stellar mass and SFR distribution. Fig. \ref{fig:structure} shows the effective stellar mass density ($\Sigma_{\rm eff}$) and the half-mass radius as a function of the stellar mass for the core, disc, and the combination of both components. We compare our measurements to the ones from \cite{Suess2019} of star-forming galaxies (SFGs) and quiescent galaxies (QGs) at $z=2.0-2.5$ and from \cite{Norris2014}, which contains data on local globular clusters (GCs), Ultra Compact Dwarfs (UCD) and compact Ellipticals (cE). In the right panel, we also plot the extrapolated growth of the half-mass radius from redshift 7.43 to redshift 1, assuming the SFR profile from Fig. \ref{fig:surface-density-radial-profiles}.

Although our $z=7.43$ galaxy has a lower stellar masses than the plotted SFGs and QGs at $z\approx2$, $\Sigma_{\rm eff}$ lies in the upper envelope of $z\approx2$ SFGs and in the range of QGs. Looking at the half-mass radius, we find that both size and stellar mass are not probed by $z\approx2$ observations. In order to facilitate a comparison and gain insight into the formation of quiescent galaxies at Cosmic Noon ($z=1-3$), we grow the observed stellar mass profile according to two simple recipes: ($i$) at a constant SFR with the observed SFR profile (green line in Fig. \ref{fig:structure}); and ($ii$) at a constant SFR but with the assumption that the scale length of the SFR surface density profile scales as $R_{\rm S} \propto \frac{1}{(1+z)}$ (motivated by inside-out growth models \cite{Mo1998}; orange line). We assume that the global SFR is constant because the decline of the star-forming main sequence and the higher SFR due to an increased stellar mass (on the star-forming main sequence) roughly cancel each other out. Although both scenarios are simplistic, both tracks naturally intersect with the $z\approx2$ size-mass relation of QGs by $z\sim1-3$, highlighting that our $z=7.43$ galaxy is a natural progenitor of the quiescent galaxy population at $z\approx2$.

We also consider how our $z=7.43$ galaxy compares to local, $z=0$ stellar systems. From the left panel of Fig. \ref{fig:structure}, we can see that this galaxy lies above $\Sigma_{\rm eff}$ of local GCs and UCD, but is comparable to low-mass ellipticals. This can also be seen from Fig. \ref{fig:compact-density-radial-profiles}, which shows the stellar mass surface density ($\Sigma_*$) as a function of radius. For comparison, we also plot the profiles of local analogues, UCDs, cE, ellipticals with a cusp, and ellipticals with a core from the compilation gathered in \cite{Hopkins2010}. The profile for \jnamespace is similar to that of cEs and more extended than that of UCDs. Interestingly, the central ($R<20~\mathrm{pc}$) density of our $z=7.43$ galaxy is within $0.2$ dex of those of massive elliptical galaxies seen in the Universe today, but we note that it contains just 0.1\% of the total stellar mass of these galaxies. Specifically, this $z=7.43$ galaxy has a stellar mass density at $R=1$ kpc of $\rm \Sigma_{*,1kpc}=6.6\;M_\odot/ kpc^2$, while the massive core ellipticals have on average $\rm \Sigma_{*,1kpc}=9.4\;M_\odot/ kpc^2$, which is a difference of 2.8 orders of magnitude. If this galaxy evolves into such a massive elliptical by $z=0$, we conclude that inside-out growth takes place in two phases: firstly as a star-forming galaxy, which we directly observe here, and then, secondly, as a quiescent galaxy from $z\approx1-3$ to $z=0$ via mergers that build up a stellar envelope \cite{Bezanson2009,Oser2012}. 

How can a $z=7.43$ galaxy build such a high central stellar mass density that is comparable to local ellipticals? Our analysis shows that the star-formation activity is dominated by the disc component. However, it is not clear whether there was an episode of disc formation prior to the peak of the SFH of the core (that is, more than 100 Myr ago): earlier disc formation is still a possibility based on the posterior distribution of the SFH (see Methods Figure \ref{fig:disc_corner}). Therefore, we speculate that the following two scenarios are possible to build up this core. The first is continuous inside-out growth, where early disc formation took place in a extremely compact disc, forming the currently observed core \cite{Mo1998}. Indeed, such compact disc-shaped objects have been observed at a redshift of more than 10 \cite{Robertson2023}. An alternative is that the disc was formed first and suffered an infall of gas into the centre due to compaction \cite{Zolotov2015, Tacchella2016}, which then formed the core. The disc would then re-form via newly accretion of gas. 

Importantly, all stellar systems, including our galaxy at $z=7.43$, are well below the maximum stellar surface density of $\Sigma_{\rm max}=10^{11.5}~M_{\odot}/\mathrm{kpc}^{2}$. This universal maximum stellar surface density of dense stellar systems is a natural consequence of feedback-regulated star-formation physics \cite{Hopkins2010, Grudic2019}.

Regarding the two inferred evolutionary tracks, we find that the constant $\Sigma_{\rm SFR}$ grows the galaxy efficiently within 1 kpc, leading to a profile at $z=2$ that overpredicts the stellar mass density of local ellipticals in the central 300 pc and is comparable to $\Sigma_{\rm max}$ at $R<10$ pc. On the other hand, the track where we scale the scale length of the SFR density has a lower $\Sigma_*$ (that is, it does not violate $\Sigma_{\rm max}$) and is more consistent with the one of massive elliptical profiles up to $R\approx3$ kpc. We conclude that the inside-out growing model is more consistent with the low-redshift ellipticals.

In summary, our finding of \jname, a core-disc galaxy with a star-forming clump during the Epoch of Reionization provides evidence for inside-out growth during the first 700 Myr of the Universe. This galaxy appears to be a potential candidate progenitor of a typical quiescent galaxy at redshift 2 and a present-day elliptical galaxy. This suggests that bulge formation can start at very early epochs and demonstrates the importance of understanding the nature of these earliest systems on spatially resolved scales.

\section{Methods}\label{Methods}

\subsection{NIRCam imaging data}

We use photometric and spectroscopic data obtained by JWST as part of the JADES \cite{Eisenstein2023} collaboration. JADES consists of the NIRCam \cite{Rieke2023} and NIRSpec \cite{Jakobson2022} Guaranteed Time Observations (GTO) instrument teams, and was established to be able to use a combination of imaging and spectroscopy to utilise the full capabilities of both instruments. We use the JADES NIRCam imaging of the Great Observatories Origins Deep Survey - South (GOODS-S) field \cite{Giavalisco2004}. This consists of imaging in F090W, F115W, F150W, and F200W short-wavelength (SW) bands, and F277W, F335M, F356W, F410M, and F444W long-wavelength (LW) bands. 

The details of the data reduction of the NIRCam data will be presented as part of the JADES programme in Tacchella et al. (in prep.), and have already been described in some detail in \cite{Robertson2023, Tacchella2023b}. Briefly, we use the JWST Calibration Pipeline v1.9.2 with the CRDS pipeline mapping (pmap) context 1039. We run Stage 1 and Stage 2 of the pipeline with the default parameters, but provided our own sky-flat for the flat-fielding. Following Stage 2, we perform several custom corrections in order to account for several features in the NIRCam images \cite{Rigby2023a}, including the 1/f noise \cite{Schlawin2020}, scattered-light effects (``wisps'') and the large-scale background. Since all of those effects are additive, we fit and subtract them.

Before constructing the final mosaic, we perform an astrometric alignment using a custom version of JWST TweakReg. We calculate both the relative and absolute astrometric correction for images grouped by visit and band by matching sources to a reference catalogue constructed from HST F814W and F160W mosaics in the GOODS-S field with astrometry tied to Gaia-EDR3 (\cite{Gaia_DR3_2021A&A...649A...1G}; G. Brammer priv. comm.). We achieve an overall good alignment with relative offsets between bands of less than 0.1 short-wavelength pixel ($<3$ mas). We then run Stage 3 of the JWST pipeline, combining all exposures of a given filter and a given visit.

\subsection{NIRSpec spectroscopic data}\label{sec:spec}

We used the NIRSpec Micro Shutter Array (MSA), with two disperser/filter combinations: PRISM/CLEAR from programme 1180 (hereafter: R100; covers the entire spectral range $0.7<\lambda<5.3$~\textmu m with spectral resolution $R=30\text{--}100$) and the high-resolution grating G395H/F290LP from programme 1286 (hereafter: R2700) to cover the region $2.9<\lambda<4.2$~\textmu m with resolution $R=2,700$.
The MSA was configured with the `three-shutter slitlet', creating an effective slit of
0.2-arcsec width and approximately 1.5-arcsec length. The exposure times were 11,292~s (R100) and 8,009~s (R2700).

For a detailed description of the data reduction, we refer to \cite{Curtis-Lake2023, Cameron2023, Curti2023b}. Here we note that we
applied wavelength-dependent path-loss corrections based on modelling \jnamespace as a point source, and extracted the spectrum from a 0.5-arcsec box.
The final reduced R100 1D and 2D spectra are shown in Fig.~\ref{fig:rgb}.

The redshift estimate is based on the [OIII]$\lambda,\lambda$4959,5007 detection in the R2700 spectrum. We obtain $z_\mathrm{spec} = 7.4303 \pm 0.0002 \text{(random)} \pm 0.0005 \text{(systematic)}$. To measure the emission-line fluxes we used the R100 data and Penalized PiXel-Fitting (pPXF) \cite{Cappellari2017,Cappellari2023}, which models simultaneously the underlying continuum, as described in \cite{Curti2023b}.

\subsection{Galaxy selection}
\label{sec:selection}

\begin{figure}
    \centering
    \includegraphics[width=0.3\columnwidth]{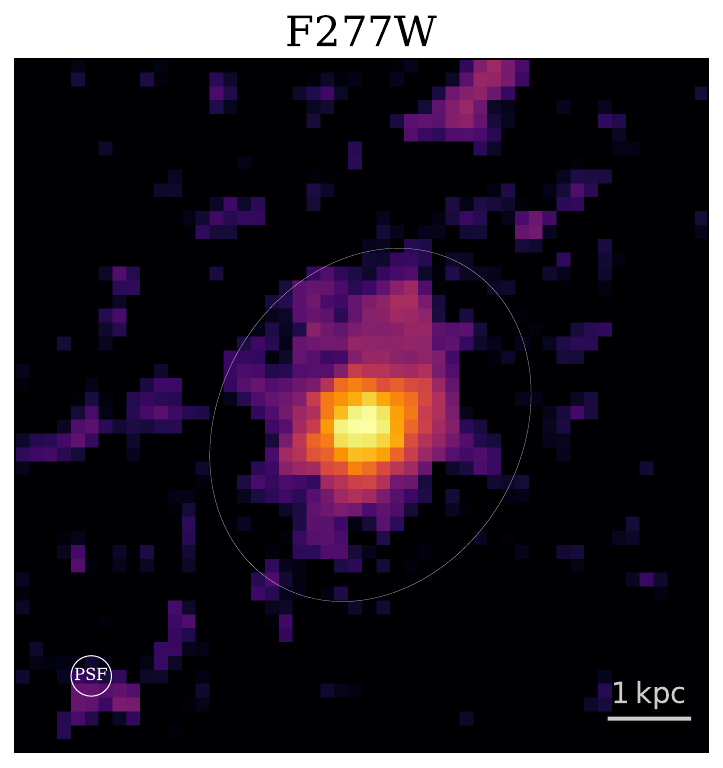}
    \includegraphics[width=0.3\columnwidth]{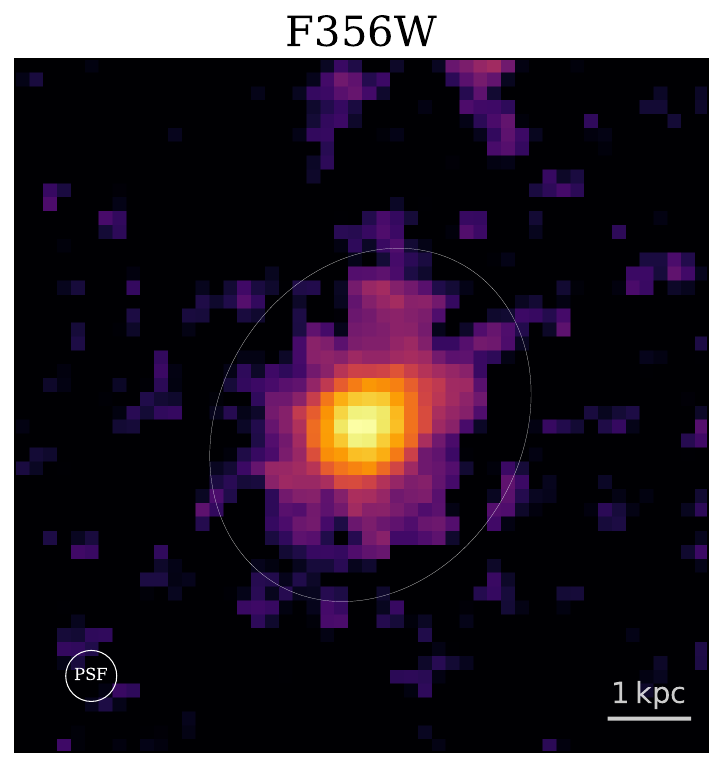}
    \includegraphics[width=0.3\columnwidth]{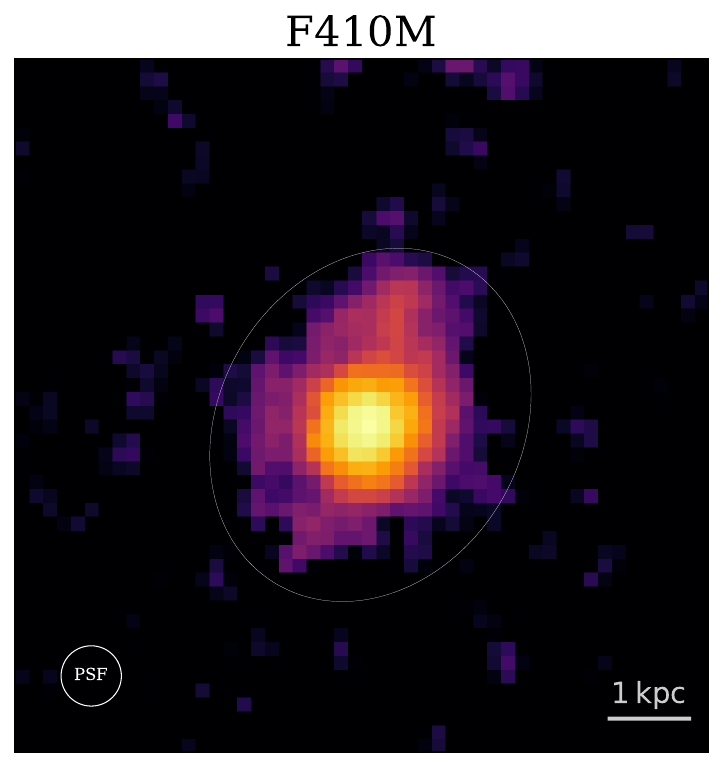}
    \includegraphics[width=0.8\columnwidth]{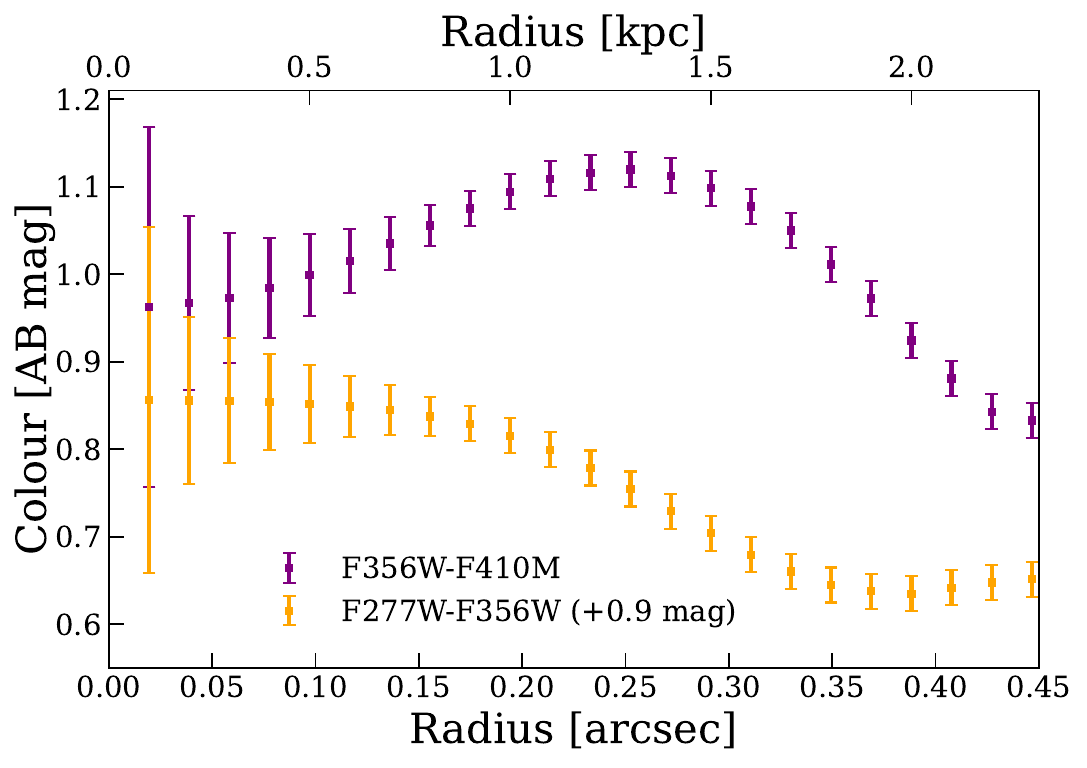}
    \caption{\textbf{Colour gradients between the centre and outskirts of the galaxy.} Images in the F277W band (upper left panel), F356W band (upper middle panel), F410M band (upper right panel) and the PSF-matched radial profiles of the F356W-F410M and F277W-F356W colour (bottom panel). The size of the FWHM of the PSF (bottom right), a bar indicating 1 kpc at $z=7.43$ (bottom left) and an elliptical aperture with a major axis length of 0.4 arcsec are shown for reference in each image cutout. The radial colour profile is computed from the PSF-matched images (PSF-matched to F444W) in order to remove gradient effects resulting from the wavelength-dependent PSF. The errors are obtained from the $1\sigma$ uncertainty map and propagated forwards.
    The F277W-F356W is offset by 0.9 mag to highlight the colour gradients seen in the profile. The colour profiles show opposite trends: the F410M excess increases while the F277W-F356W colour (tracing the rest-4000\AA\/ spectral region) gets bluer toward the outskirts. This indicates that the outskirts -- dominated by the disc -- has younger stellar population than the central region.}
    \label{fig:image_color}
\end{figure}

\begin{figure}
    \centering
    \includegraphics[width=0.8
    \columnwidth]{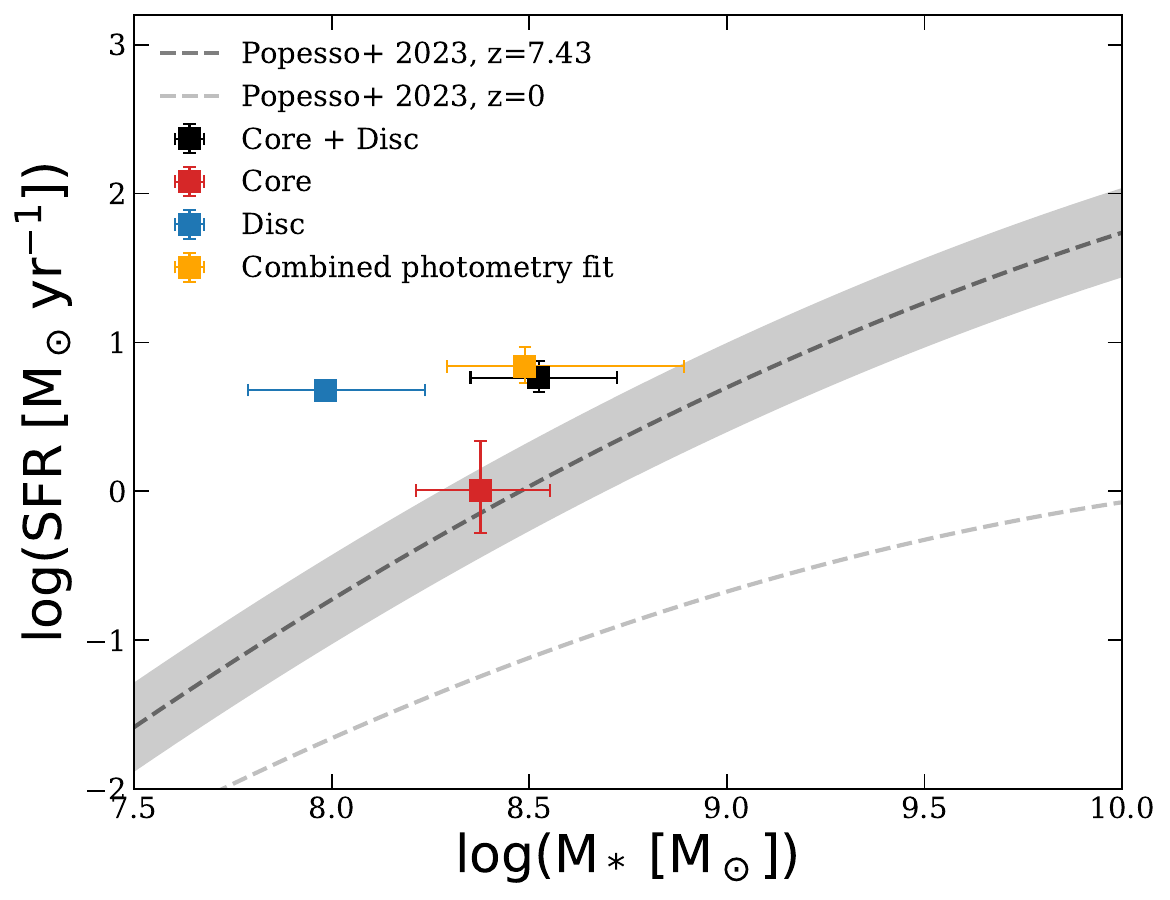}
    \caption{\textbf{Star-formation activity relative to the star-forming main sequence.} The SFR (averaged over 10 Myr) against the stellar mass for the core, disc and combined components plotted in red, blue and black. The result of fitting the core+disc+clump photometry in also plotted in orange. The data is presented as the median of the distribution with the errors corresponding to the 16th and 84th percentiles.
    For reference, the star-forming main sequence (SFMS) \cite{Popesso2023} extrapolated to redshift 7.43 and at 0 is plotted as black and grey dashed lines, respectively (the shaded region corresponds to a 0.3 dex uncertainty region on the SFMS at $z=7.43$). All components show elevated SFRs compared to the SFMS (by 0.6 and 1.2 dex for the core and disc, respectively).}
    \label{fig:sfms}
\end{figure}

In order to model the spatially resolved stellar populations, we selected \jnamespace from the JADES GOODS-S imaging region with F335M and F410M medium-band coverage and a spectroscopic redshift. We used spectroscopic redshifts from both JADES and FRESCO \cite{Oesch2023}, focusing on the redshift interval $z=7.0-7.8$, as we wanted to both probe the very earliest galaxies while also having the Balmer break falling within our filters. Out of these ($\approx$20) galaxies, \jnamespace appeared to be the most intriguing with evidence of a core-disc structure and colour gradient (see Fig. \ref{fig:image_color}).  We also note that this galaxy was included in the six sources analysed in \cite{DeGraaff2023} where they found evidence for rotation likely tracing the disc component.

Due to this selection procedure, we make no claims about population statistics for this type of galaxy at these redshifts, only that it is a fascinating object in its own right. In a future work, we will explore the population statistics for a mass-complete sample of similar (bulge-disc) galaxies. Importantly, this galaxy does not seem to be peculiar given its stellar mass and redshift: it is only slightly above the extrapolated star-forming main sequence (\cite{Brinchmann2004, Noeske2007, Speagle2014, Baker2023a, Popesso2023}; see Fig. \ref{fig:sfms}) and shows the typical emission line properties for galaxies at this redshift (refer to Sections \ref{sec:em_lines} and \ref{sec:SF_AGN}). \jnamespace has previously been identified in HST imaging for the GOODS-S field as a Lyman break galaxy at $z\sim7-8$ (it is the source UDF-3244-4727 in \cite{Bouwens2008} based on HST-NICMOS and ACS imaging, and was independently selected in subsequent HST/WFC3 imaging as GS.D-YD4 in \cite{Lorenzoni2013}).

Fig. \ref{fig:image_color} shows images of the galaxy in the F277W band (upper left panel), the F356W band (upper middle panel), the F410M band (upper right panel), and the PSF-matched radial profiles of the F356W-F410M and F277W-F356W colour (bottom panel). The radial colour profile is computed from the PSF-matched images (PSF-matched to F444W) in order to remove gradient effects resulting from the wavelength-dependent PSF. Interestingly, we find opposite trends for the two colours: the F356W-F410M colour gets redder toward the outskirts, while the F277W-F356W colour gets bluer. Since the F356W-F410M colour traces mainly the emission line excess ([OIII] and H$\beta$ lines), we find that those emission lines become more prominent towards the outskirts. On the other hand, the F277W-F356W colour traces the rest-frame 4000\AA\/ spectral region, implying that the centre shows a Balmer/4000\AA-break, while the outskirts show a Balmer jump. This indicates that the outskirts have a younger stellar population than the central region. Interestingly, the ionised gas of \jnamespace is actually consistent with a rotationally supported component with $v(r_{\rm eff})/\sigma_0\approx1.3$, as recently shown in \cite{DeGraaff2023}.

\subsection{Morphology and photometry with ForcePho}
\label{sec:forcepho}

\begin{figure}
    \centering
    \includegraphics[width=\columnwidth]{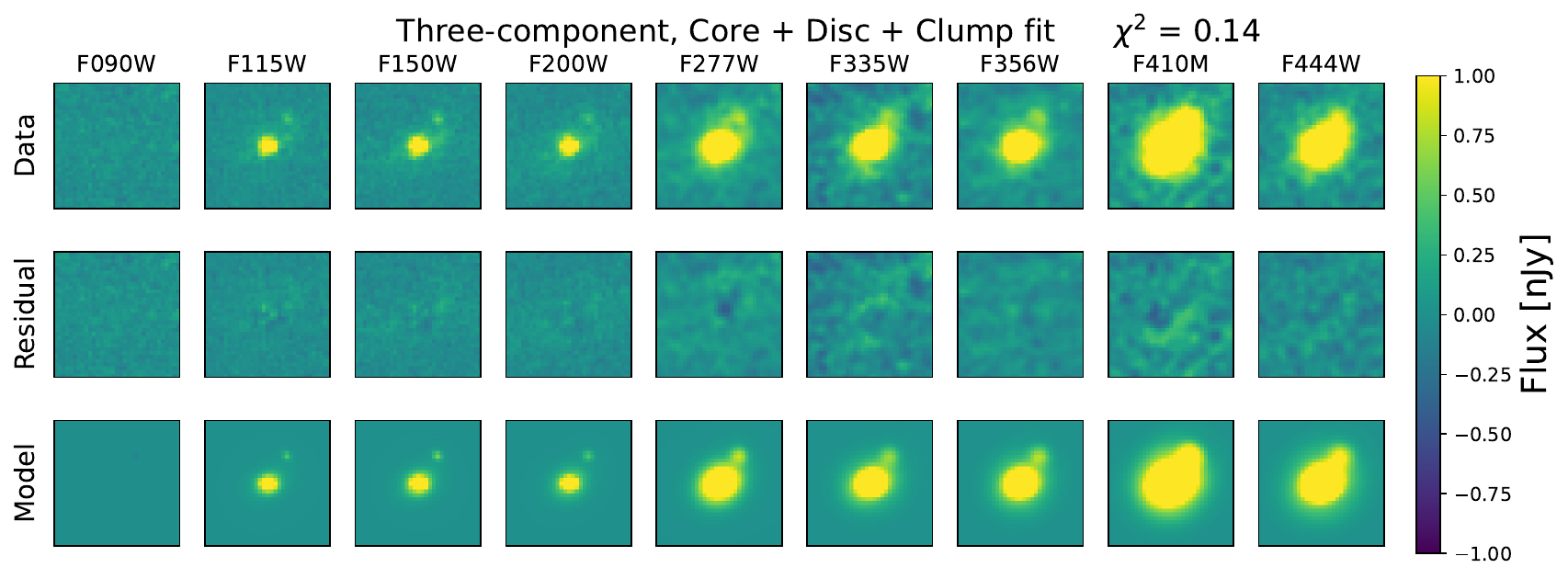}
    \includegraphics[width=\columnwidth]{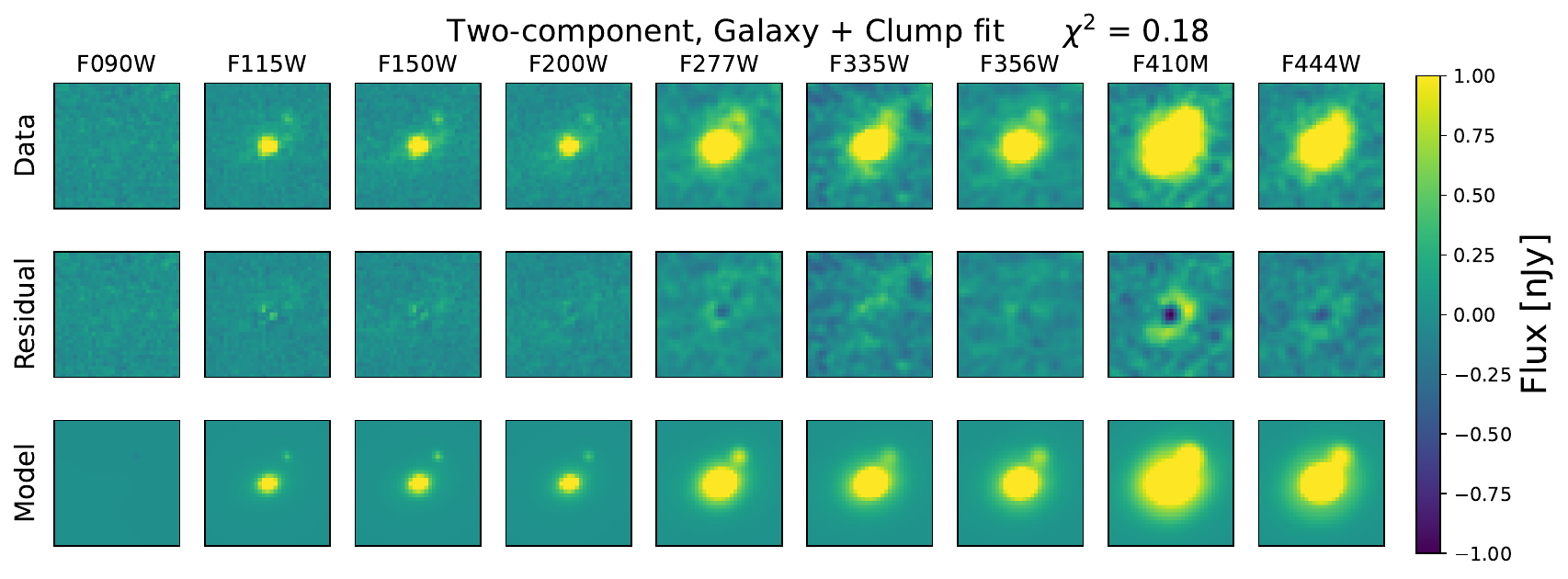}
    \includegraphics[width=\columnwidth]{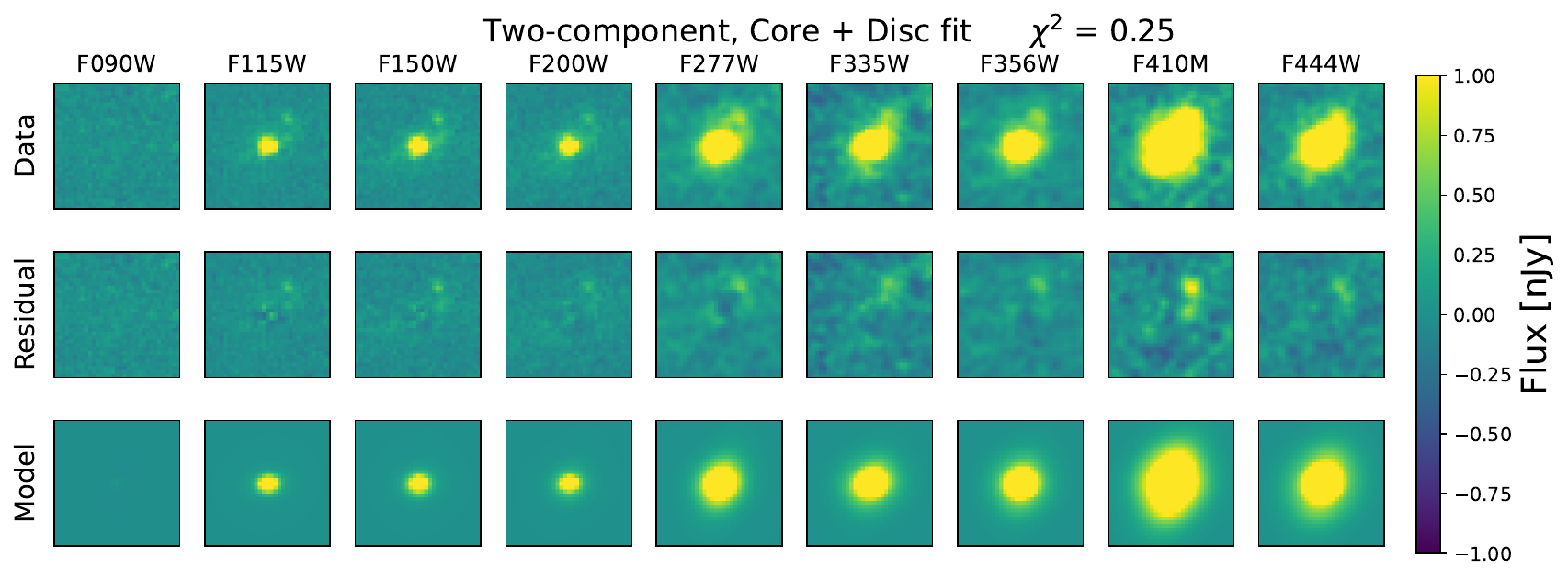}
    \includegraphics[width=\columnwidth]{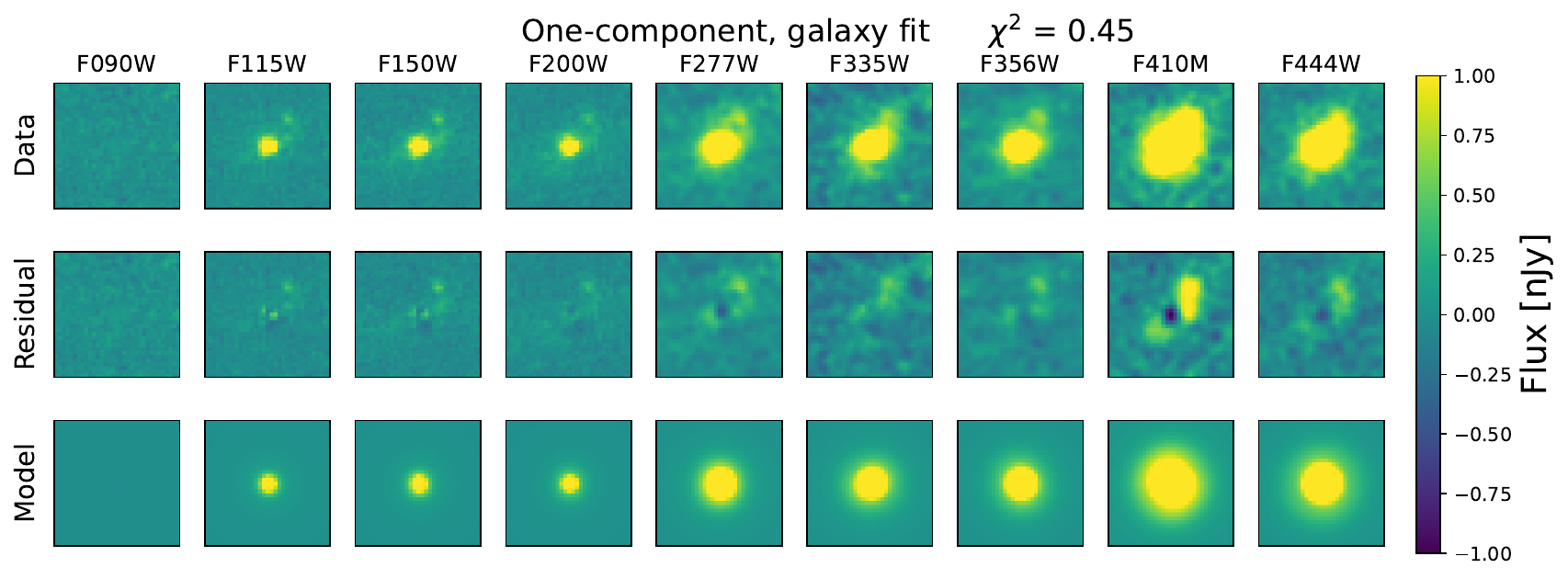}
    \caption{\textbf{Data, model, and residual maps for various ForcePho models.} For each of the four main setups we plot the data, residual and model. The four setups are our fiducial, three-component fit (top panels), a two-component fit consisting of galaxy+clump (upper middle panels), a two-component fit consisting of a core+disc (with no clump),and a one-component single galaxy fit (bottom panels). It shows that the three-component fit appears to reproduce the data significantly better, as is quantified through the $\chi^2$-value for the fit. This can also be seen in the residual figures for the other fits which all show more significant residuals (alongside their higher $\chi^2$ values).}
    \label{fig:data-model-rsidual-comparison}
\end{figure}

It is challenging to assess the morphology and spatially resolved stellar populations of galaxies at $z>3$ because ($i$) they are compact with typical sizes of $0.1-0.4$ arcsec, i.e., of the order of size of the NIRCam/F200W PSF, and ($ii$) the NIRCam PSF FWHM varies by a factor 4 from the bluest (F115W) to the reddest (F444W) band. Given this variation in the PSF FWHM, there are two routes to analysing the data: (1) performing pixel-by-pixel SED fitting on images convolved to the F444W resolution, or (2) modelling the galaxy's deconvolved light distribution in each band independently. Convolving to F444W resolution would lose the high spatial resolution available to us in the blue bands, which provides crucial insights into the morphology of the source; therefore, we choose to forward-model the galaxy's light distribution. Importantly, the observed colour gradient is not caused by the PSF as we show in Fig. \ref{fig:image_color} after homogenising for the PSF, the colour gradient is still present. This implies that there is an intrinsic colour gradient. In order to model the intrinsic colour gradient, we have to vary the structural components of the galaxies as a function of wavelength (just changing the normalisation will not be enough because the radial component will cancel out in the colour term). This can be done via: ($i$) fitting an individual S\'ersic profile with wavelength-dependent size and/or S\'ersic index; or ($ii$) fitting a two-component S\'ersic model with fixed size and S\'ersic index, but wavelength dependent normalisation. We chose option ($ii$) because it has a stronger physical motivation. A given physical component (bulge, disc, or clump) has a certain morphology (size and shape) and stellar population properties (stellar mass, SFH, etc.). This means that a component has a certain SED, that is, the flux changes as a function of wavelength, but the shape remains the same. Furthermore, fixing the shape parameters will allow the use of the full signal-to-noise ratio across all bands.

A challenge with this approach is that we have to choose a parametric model that describes the galaxy well. As outlined below (Section~\ref{sec:1v2}), we experimented with different models, finding that two central components with a clump in the outskirts are able to reproduce the light distribution of \jnamespace in all bands the best (both visually and quantitatively, see Fig. \ref{fig:data-model-rsidual-comparison}) . Our main aim here is to describe the stellar populations on spatially resolved scales (see Fig.~\ref{fig:structure}), while the physical interpretation of the different components is of secondary importance, but remains of interest given the insights into bulge-disc systems at lower redshifts \cite{Simard2011, Tacchella2015, Lang2014}.

We choose here to forward model the light distribution in all 9 NIRCam images using ForcePho (Johnson et al., in prep.). ForcePho fits multiple PSF-convolved S\'ersic profiles simultaneously to all individual exposures and filters by sampling the joint posterior distribution via Markov Chain Monte Carlo (MCMC). This allows us to take into account and measure the covariances between all the parameters. We run ForcePho on the individual NIRCam exposures, which is a key advantage over other codes that run on mosaics, such as \texttt{galfit} \cite{GALFIT_2010AJ....139.2097P}, \texttt{Lenstronomy} \cite{Lenstronomy_2018PDU....22..189B} or \texttt{ProFit} \cite{Robotham2017}. Firstly, when the individual cal frame images (stage 2 products) are co-added to build the final mosaic, information is lost by construction. Working therefore on the mosaic means working on data with less information than the full set of individual exposures. This is particularly important for compact objects, such as \jname. The individual exposures capture these compact objects with several different pixelizations (thanks to different dither positions), while the mosaics are a single-pixel representation. Information is also lost about the correlation between pixel fluxes in the mosaics. Secondly, alternative methods work with empirical PSFs (ePSFs), which are based on a few stars that are not saturated, leading to significant uncertainty in the outskirts of the ePSFs due to noisy outskirts of individual stars. Furthermore, the ePSFs are only marginally oversampled, which leads to uncertainties in the convolution. The PSFs of the cal frame images can be well described with WebbPSF (\cite{Rigby2023, Perrin2014}; see also Section \ref{sec:psf_approx}). Therefore, tools such as ForcePho that can work on individual exposures have a significant advantage over tools that work on mosaics with ePSFs. Furthermore, ForcePho has been successfully applied to modelling multiple components in \cite{Tacchella2023a, Robertson2023, Tacchella2023b}. Since ForcePho needs to perform many convolutions, it works with Gaussian mixture models for both the PSF and the S\'ersic profile. The PSF is approximated with 4 positive Gaussians. The introduced systematics are investigated in Section~\ref{sec:psf_approx} and motivate us to assume an error floor for the parameters estimated by ForcePho ($20\%$ for the effective radii and 0.3 for the S\'ersic index of the core).

We run ForcePho assuming a three-component model: two central components and one off-centred clump. Our data prefer this three-component model over a more simple model (one- or two-component model), as shown in Section~\ref{sec:1v2} both visually and quantitatively. 
We assume that the structural parameters are constrained by a combination of the bands, while the flux is fit individually in each band. For the two central components, we fit for the centre, the axis ratio, the position angle, and the size. The prior on the size is uniform from 0.001'' to 1.0''. Importantly, we fix the S\'ersic index $n$ of one of the central components to 1.0 (sampling it from a range of 0.99 to 1.01), while the other central component is allowed to vary between $2<n<5$. The motivation for fixing the S\'ersic index to 1 for one of the central components comes from restricting the number of degrees of freedom and from lower-redshift observations \cite{Simard2011, Tacchella2015}, where so-called bulge-disc decomposition fits have been shown to describe well the light and stellar population distributions. The assumption of an exponential disc ($n=1$) is theoretically motivated as gas that cools in a halo leads to an exponential star-forming disc \cite{Mo1998}, while observed discs are well fit by an exponential profile \cite{Nelson2016, Belfiore2018}, hence most bulge-disc decomposition studies fix the S\'ersic index to 1 \cite{Kormendy2004,Simard2011,Lang2014}. Furthermore, specifically for this galaxy, \cite{DeGraaff2023} shows evidence for rotation in it.
Therefore, we call the $n=1$ component a ``disc'', while the second smaller component with $n=2-5$ is referred to as a ``core''. We fit the off-centred clump as a quasi-point source whose radius is fixed to a maximum of 0.01 arcsec (51 pc) and a fixed S\'ersic index of 1 to suppress prominent wings.  
In total, we fit a model with 43 free parameters and two fixed parameters (the S\'ersic indices of the disc and the clump).  

\begin{figure*}
\begin{picture}(200,300)
\put(0,0){\includegraphics[scale=.5]{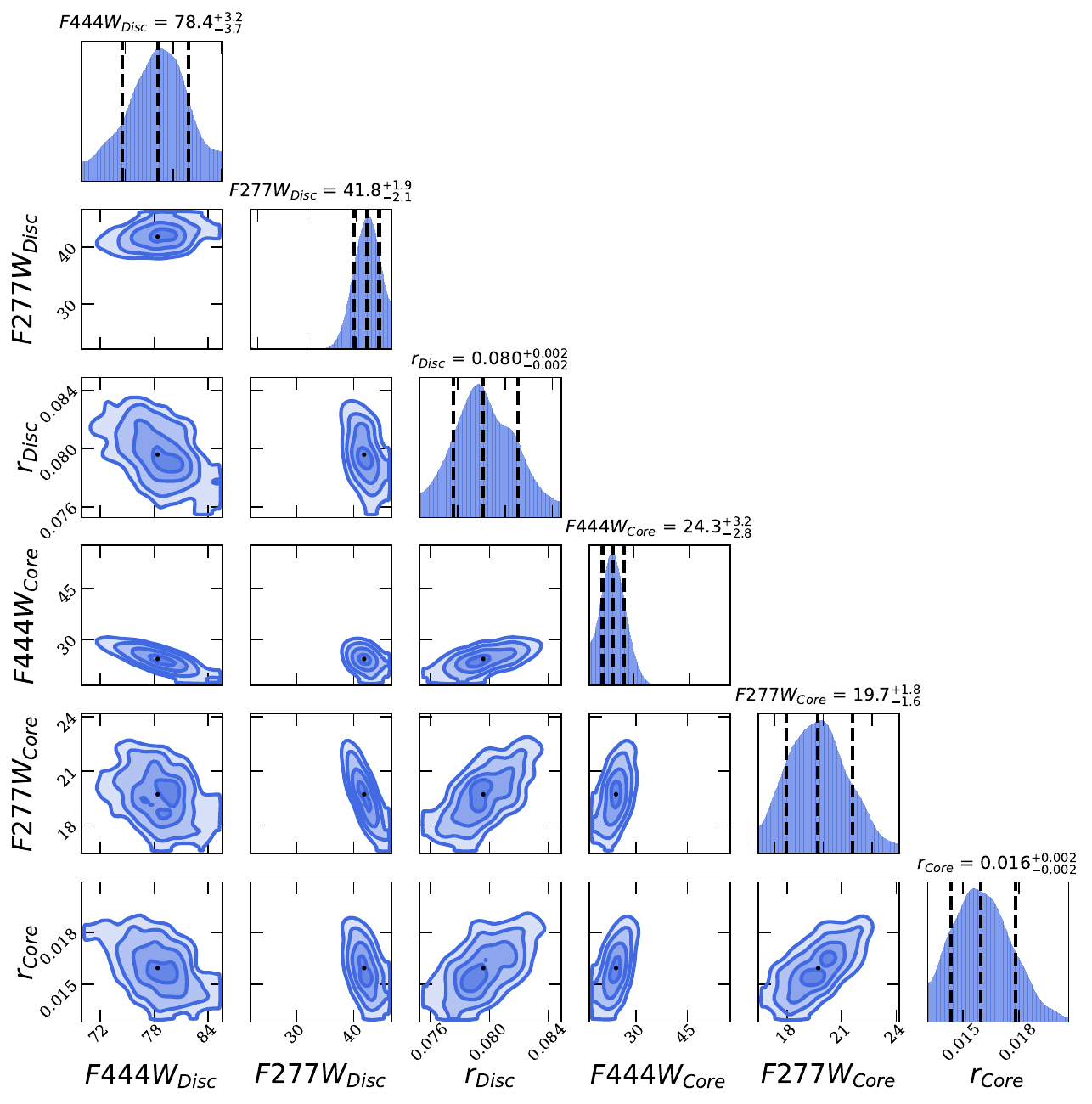}}
\put(165,170){\includegraphics[scale=.3]{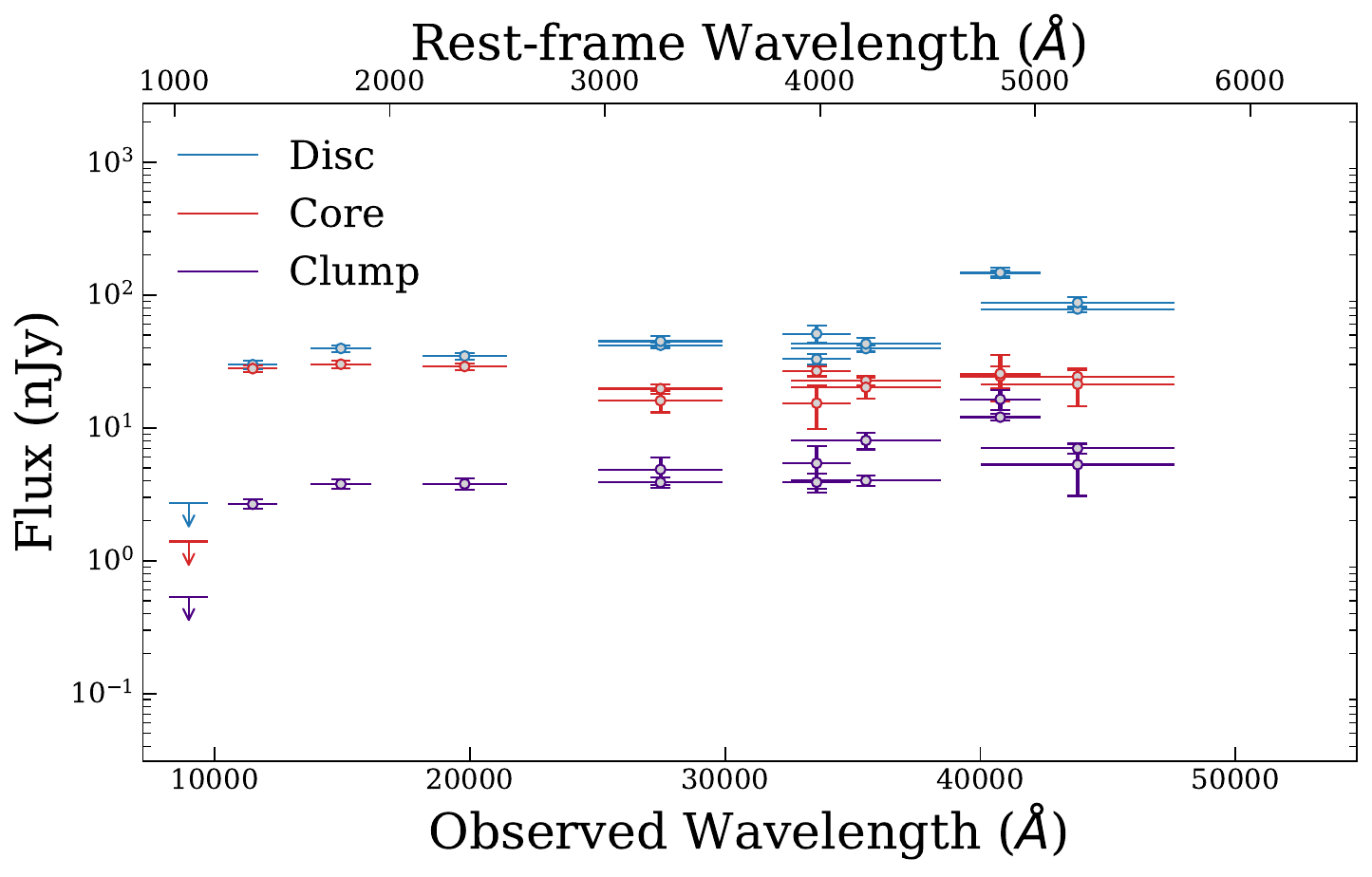}}
\end{picture}
\caption{\textbf{Posterior distributions for the ForcePho fit and spectral energy distribution for the resulting components.} Left panel: corner plot of the posterior distributions for flux in the F444W filter, F277W filter, and half-light radius for the disc and core component from the ForcePho fit. The corner plot shows that ForcePho constrains these parameters well. We find that the core and disc have a size of $16\pm2$ mas (82 pc) and $80\pm2$ mas (412 pc), respectively. We boost the errors of the core and disc to be a minimum of 20\% in order to better account for systematics. Right panel: the spectral energy distribution (SED) of the core, disc and clump components. The x errors correspond to the filter widths, and the y errors correspond to the 1$\sigma$ uncertainty propagated forwards from the error map.}
\label{fig:forcepho}
\end{figure*}

We check the success of our fits by exploring the overall data, residual, and model images (see Fig. \ref{fig:data-model-rsidual-comparison}, upper plot). Our fit's residuals are generally consistent with the background as can be seen from the imaging, with slight excess residuals seen in the F410M filter likely from the strong emission lines.  We find that the best-fit centre of the core and disc align well, with an offset of 0.005'', which is less than 0.2 SW pixel and less than the size of either central components. The core has a small effective size of $16\pm3$ mas and a S\'ersic index of $2.0\pm0.4$, while the disc component has a larger size with $80\pm16$ mas. The left panel of Fig. \ref{fig:forcepho} shows the posterior distributions of key parameters from the ForcePho fit (the flux in the F444W and F277W bands and the half-light radius). As can be seen from the corner plot in Fig. \ref{fig:forcepho}, ForcePho obtained informative posterior distributions for both the central core and the disc components. The MCMC approach of ForcePho also allows us to assess the degeneracies in the fitting, as apparent from the covariance in the core and disc fluxes of F444W and F277W. Importantly, the posterior distributions only include the statistical uncertainties and not systematic effects, and we assume an error floor for both the fitted photometry (of 5\%) and for the morphological parameters (20\% for the size, and 0.3 for the S\'ersic index of the core). This is motivated by the tests described below. In addition, the right panel of Fig. \ref{fig:forcepho} shows the ForcePho spectral energy distributions (SEDs) of the core, disc and clump components, which indicate different stellar populations for the three different components.

We also test leaving the S\'ersic index of the disc free to vary from 0.7-1.3 allowing for a broader range of values. We find that we obtain a disc with half-light radii of 0.12'' ($\pm 0.02$) where the S\'ersic index becomes 0.75 ($\pm 0.30$). Although this is broadly consistent with our fiducial run ($2\sigma$), we note that this run is close to the prior boundary, which cannot be extended to lower S\'ersic indices due to the Gaussian mixture approach of ForcePho. We acknowledge that a lower S\'ersic index might be preferred by the data. Because of the degeneracy of S\'ersic index and size, this run also implies larger sizes, which would only strengthen our results, where the star-forming disc is more extended than the core of the galaxy. Finally, we check the chi-squared value of this run, finding a value of 0.17 and a reduced chi-squared value of 8.5. This is larger than our fiducial fixed S\'ersic component run, which is still the preferred model and includes fewer degrees of freedom.

As shown in Section~\ref{sec:1v2}, both visually and quantitatively, two central components reproduce the observed light distribution of \jname. But what is the evidence for calling the two central components ``disc'' and ``core''? Firstly, focusing on the structure, we find that the effective size of the disc is over 4 times larger than the core, which indicates that the core is a much more compact component than the disc. The S\'ersic index of the core is consistent with a ``pseudo-bulge'' component ($2.0\pm0.4$), that is, we do not find any evidence for a classical bulge-like component. Importantly, we stress that our disc and core are photometric components, and we cannot say anything about the kinematics of those components from our analysis. However, as mentioned above, the ionised gas of \jnamespace is actually consistent with a rotationally supported component with $v(r_{\rm eff})/\sigma_0\approx1.3$, as recently shown in \cite{DeGraaff2023}, which motivates us to refer to the extended component as a ``disc''. The resulting SEDs of the core and disc components are shown in the right panel of Fig.~\ref{fig:forcepho}. These SEDs lead to different stellar populations for the core and the disc (see Section \ref{sec:SEDFitting}), consistent with the idea of a slightly older core and a younger disc component. In summary, on the basis of the structure and the inferred SEDs and stellar populations, we find support for interpreting the two central components as a disc and a core. 

We find a consistent interpretation from the direct colour analysis presented in Section~\ref{sec:selection} and Fig.~\ref{fig:image_color}. The F277W-F356W and the F356W-F410M colour profiles indicate an outskirt that is dominated by younger stellar populations than the central region. A direct comparison with the core and disc colour obtained from ForcePho should be taken with a grain of salt, because our decomposition allows for mixing of the different components at fixed radius. We perform aperture photometry on the PSF-matched mosaics using a central aperture of 0.2'' and an outskirt aperture of 0.4''. We find that the colours for the centre are similar to that of the core although with larger differences in the medium bands likely stemming from emission lines (F277W-F356W=$\rm 0.15_{-0.07}^{+0.07}\;mag$, F356W-F410M=$\rm 0.09_{-0.17}^{+0.30}\;mag$ for the core and F277W-F356W=$\rm -0.06\pm0.06\;mag$, F356W-F410M=$\rm 1.03\pm0.05\;mag$ for the centre), whilst the colours for the outskirts are similar to that of the disc (F277W-F356W=$\rm -0.06_{-0.06}^{+0.06}\;mag$, F356W-F410M=$\rm 1.41_{-0.05}^{+0.11}\;mag$ for the disc and F277W-F356W=$\rm -0.19\pm0.14\;mag$, F356W-F410M=$\rm 1.07\pm0.14\;mag$ for the outskirts).

\subsection{Motivation for the multiple component fit}
\label{sec:1v2}

It is crucial to check whether a multi-component fit is warranted by the NIRCam imaging data. Part of our motivation for a multiple-component fit is due to an expectation that the normalisation of different components will vary as a function of wavelength, therefore the structure will also vary with wavelength. One example of this is to fit the short-wavelength (SW) and long-wavelength (LW) bands separately. We find that the LW bands fit is broadly consistent with the fiducial run ($r_{half}$=0.028'') for the core and 0.088'' for the disc, whilst the SW fit inverts the radii of the core and disc components 0.042’’ and 0.029’’ respectively. We would expect the LW data to best mirror the fiducial run as we have more possible exposures in this wavelength range, while for the SW bands we are relying on only the exposures from four filters, hence the obvious degeneracies between the components.
Fig. \ref{fig:data-model-rsidual-comparison} upper panel shows the data, the residual, and the best-fit model for the three-component fit in all the JADES filters. We see that we can reproduce the observed light distribution in each filter for this three-component model. We then compare this with two simpler models. We incorporate a $\chi^2$ measure for the ForcePho fits, where we measure the chi values of all pixels within a segmentation map image of the galaxy, as calculated from the F444W image. We then calculate the $\chi^2$ value as the sum of the residuals of all stacked pixels within the segmentation map divided by the noise (estimated from the background). The reason we use a segmentation map is twofold: firstly, we do not want the metric to be dependent upon the size of the cutout, secondly, if we included all pixels in the cutout we would be dominated by the background pixels rather than pixels within the galaxy. We find that for the three-component model, this gives us a $\chi^2$ value of 0.14. We note that the exact values depend on the exact choice of the segmentation map.

We run a single-component fit as a comparison to the three-component model. This means that we treat the whole galaxy as a single component and allow its S\'ersic index to vary freely from 0.8-6, enabling it to be modelled as a disc or bulge-like component.  As can be seen in Fig. \ref{fig:data-model-rsidual-comparison} bottom, we find that the 1-component fit (bottom panel) under subtracts the central region (i.e., the core) compared to the three-component fit (top panel); it also significantly fails to fit the clump. The single component fit gives a $\chi^2$ value of 0.45 indicating that it has struggled more than the three-component run to reproduce the observed light distribution.
Our next test is to run a single component fit for the main galaxy plus the clump (to test whether the core disc fit is warranted). Once again, the model fails to account for the flux in both the centre and the outskirts, as seen in the upper middle panel in Fig. \ref{fig:data-model-rsidual-comparison} compared to the three-component fit in the top panel. This fit gives a $\chi^2$ value of 0.18, showing an improvement over the single-component fit, but it still does not reproduce the observed light distribution as well as the three-component fit.

We also test the effect of ignoring the clump by fitting for just the core and disc components. The most significant change is that this extends the half-light radius of the disc component from 80 mas to 129 mas, whilst increasing the core radius to 0.039 mas; in essence, the components expand in radii to try to fit light from the clump. The S\'ersic index of the core becomes 2.31. Importantly, we find consistent fluxes for the core and disc components compared to the three-component analysis: changes are well within the uncertainties. Specifically, the core fluxes change less than $10\%$ in all bands except F410M and F444W, for which the fluxes increase by $26\%$ and $25\%$, respectively. Since the disc fluxes for those bands remain unchanged (changes of less than $3\%$), this indicates that the core picks up clump long-wavelength light. We find a $\chi^2$ value of 0.25 for the fit showing the importance of modelling the clump component.
In summary, this test shows that our main results regarding the stellar population differences between disc and core still hold when ignoring the off-centred clump. 

In addition to the $\chi^2$-values mentioned above, we also test metrics for goodness of fit that can incorporate the number of degrees of freedom. This is an important consideration, as the $\chi^2$ value alone, while informing us which model best reproduces the data, cannot tell us whether it is worth the increased complexity of the model. However, determining a reduced chi-squared is not trivial in this case because ForcePho does not optimise the total chi-square itself, but rather fits the individual exposures. As with the aforementioned chi-squared, we only consider the pixels within the segmentation map. Our reduced chi-squared is the previously measured chi-squared multiplied by the degrees of freedom. We define the degrees of freedom as the number of independent pixels within the segmentation map minus the number of parameters of the model. The number of independent pixels within the segmentation map is selected as the total number of pixels in the map minus the number of pixels within the full-width half-maximum of the PSF. The number of model parameters are the free parameters within the fitting for the different component fits. For the three-component core+disc+clump fit, we obtain a reduced chi-squared value of 6.7. For the two-component galaxy+clump fit we obtain a value of 11.4. The two-component core+disc fit gives a value of 15.7 and the single-component galaxy gives 35.2. In summary, even after considering the degrees of freedom, our fiducial three-component core+disc+clump has the smallest reduced chi-square value.

\subsection{PSF approximations in ForcePho}
\label{sec:psf_approx}

ForcePho approximates the JWST PSFs with a Gaussian mixture model. Four Gaussians are able to describe the key components of the JWST/NIRCam PSFs as provided by WebbPSF. In this section, we explore the validity of these PSF approximations specifically for the data and \jnamespace presented in this work.

\begin{figure}
    \centering
    \includegraphics[width=0.47\columnwidth]{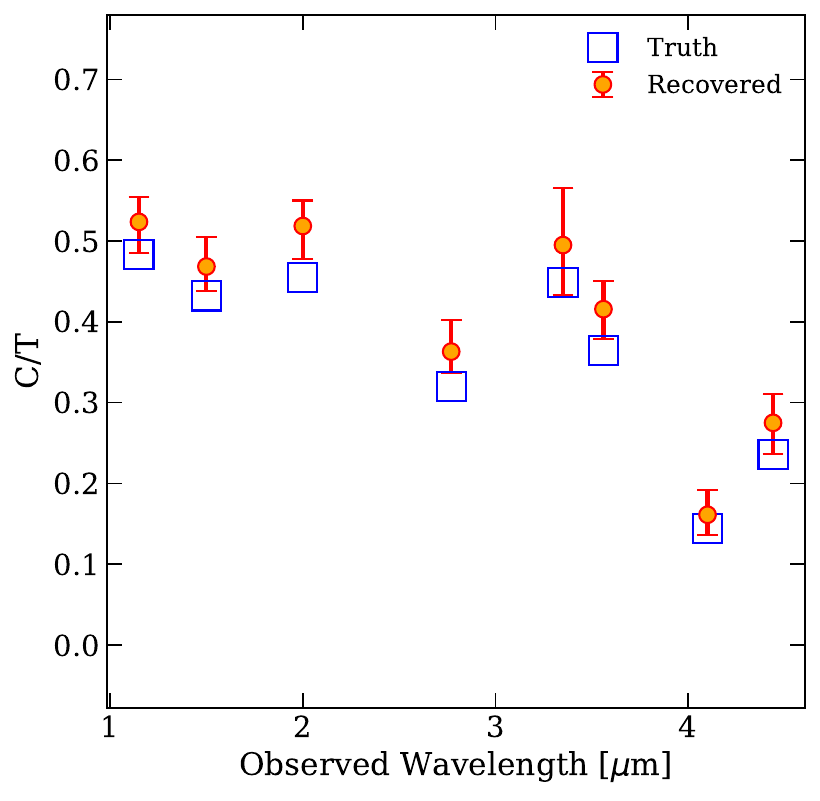}
    \includegraphics[width=0.49\columnwidth]{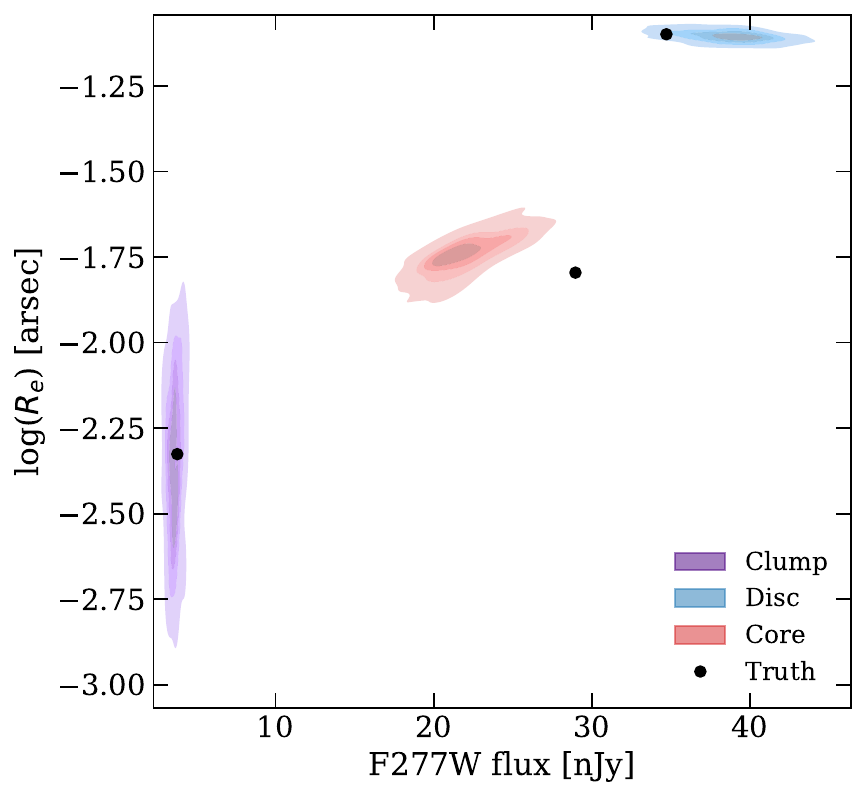}
    \caption{\textbf{Component recovery test with ForcePho.} We simulate the best-fit 3-component model with Galsim, assuming the PSF from WebbPSF. We then refit this scene with ForcePho to determine the validity of ForcePho's PSF approximations. Left: the core-to-total (C/T) ratio from the original fit (truth) and the fit (recovered), plotted against filter wavelength. The data is presented as the median of the distribution with the errors corresponding to the 16th and 84th percentiles.
    Right: the half-light radius against flux in the F277W filter, the contours correspond to the posterior for the recovery tests, while the black point is the value from the original fit. These two plots show that ForcePho's PSF approximations are close to the true value and that we recover the fitted values close to the errors. The deviation seen in the contour plot is driven by the compact half-light radius of the core.}
    \label{fig:psf_recovery test}
\end{figure}

We simulate the best-fit 3-component model (Section \ref{sec:forcepho}) with Galsim \cite{Galsim_2015}. Specifically, we produce the full set of Stage 2 products as given by our observations, assuming the best-fit 3-component model for our galaxy and PSFs directly obtained from WebbPSF. We then refit with a 3-component model with ForcePho, assuming the same setup as described above. Figure \ref{fig:psf_recovery test} shows the recovery of the core-to-total (C/T) ratio as a function of the filter wavelength (left panel) and the half-light radius versus the F277W flux for the three components (right panel), with the contours corresponding to the posteriors and the black point is the input value. We are able to recover to within $1\sigma$ the C/T ratio, but all wavelengths are biased high, i.e. the recovered cores are overestimated relative to the discs. We find that the C/T ratios at all wavelengths are biased high, i.e. the recovered cores are overestimated relative to the discs. Although we can recover the C / T ratios within $1\sigma$, this clearly shows a bias on the level of $5-10\%$. It is comforting that this bias is nearly wavelength independent (variations within $<5\%$), which implies that only the normalisation of the SED of the core and the disc are affected, but not their colours. This in turn means that we possibly overestimate the stellar mass of the core by $5\%$ relative to the disc. The right-hand plot of Figure \ref{fig:psf_recovery test} shows that we are able to recover the morphological parameters (i.e. half-light size) well, within 0.05 dex ($10-20\%$). In order to account for possible systematics, we assume an error floor of 20\% for the measured sizes, i.e., inflate the uncertainties if necessary. 

This test leaves open whether WebbPSF provides accurate PSFs. We have tested this in detail in \cite{Ji2023} by constructing and comparing empirical PSFs both from true observed stars (called empirical PSFs [ePSFs]) and from WebbPSF point sources injected at the Stage 2 level images (called model PSFs [mPSFs]). Specifically, we construct ePSFs using the empirical method proposed by \cite{Anderson2000}. This method solves the centroids and fluxes of a list of input point sources, and then stacks all the point sources together to get the ePSF. For the list of point sources, we visually identified a sample of 15 isolated point sources from JADES, which are bright but unsaturated. We obtain mPSFs by injecting WebbPSF-based PSFs into the Stage 2 images and then mosaiking them in the same way as our normal mosaics, producing ``star'' images. We have then constructed mPSFs from those star images in the same way as the ePSFs. Importantly, we find excellent agreement between ePSF and mPSF, with a typical difference $\lesssim 1\%$ in  the radial profile of enclosed energy, from the central pixel out to 3 arcsec. This implies that the prediction of {\sc WebbPSF} is accurate.

\subsection{SED fitting with Prospector}
\label{sec:SEDFitting}

Prospector \cite{Prospector2021} is an SED fitting code which takes in photometric fluxes and flux errors and fits model SEDs to them. It uses the Dynamic Nested Sampling package Dynesty \cite{Speagle2020} and models the stellar populations via Flexible Stellar Population Synthesis (FSPS \cite{Conroy2009, Conroy2010}), where we use MIST isochrones \cite{Choi2016} and a Chabrier \cite{Chabrier2003} initial mass function. 

We assume a stellar population model similar to that in \cite{Tacchella2022}. Briefly, we assume a flexible SFH with 6 different time bins, where the most recent bin covers the last 5 Myr, with the other bins being split between 5 Myr and 520 Myr ($z=20$) in log steps. We use the standard continuity prior \cite{Leja2019}, which weights against a bursty SFH. We use a top-hat prior on the log stellar mass, where it varies from 6 to 12. To model the effect of dust attenuation we use a flexible two-component dust model \cite{Charlot2000}, which models a separate birth cloud component (attenuating emission from gas and stars formed in the last 10 Myr) and a diffuse component (attenuating all emission from the galaxy). We use a joint prior on the ratio between the two dust components, where the prior is a clipped normal between 0 and 2, with a mean of 1.0 and a standard deviation of 0.3. The prior on $\tau_V$, the optical depth of the diffuse component in the V band, is a clipped normal ranging from 0 to 4 with a mean of 0.3 and a standard deviation of 1. The slope of the dust attenuation law of the diffuse component is a free parameter and is modelled as a power law multiplication of the standard \cite{Calzetti2000} law (with a top-hat prior from -1 to 0.4). We also use a top-hat prior for the log stellar metallicity with a minimum of -2.0 and a maximum of 0.19. For the nebular component, managed by FSPS, we have a freely varying ionisation parameter and gas-phase metallicity \cite{Byler2017}.

\begin{figure}
    \centering
    \includegraphics[width=\columnwidth]{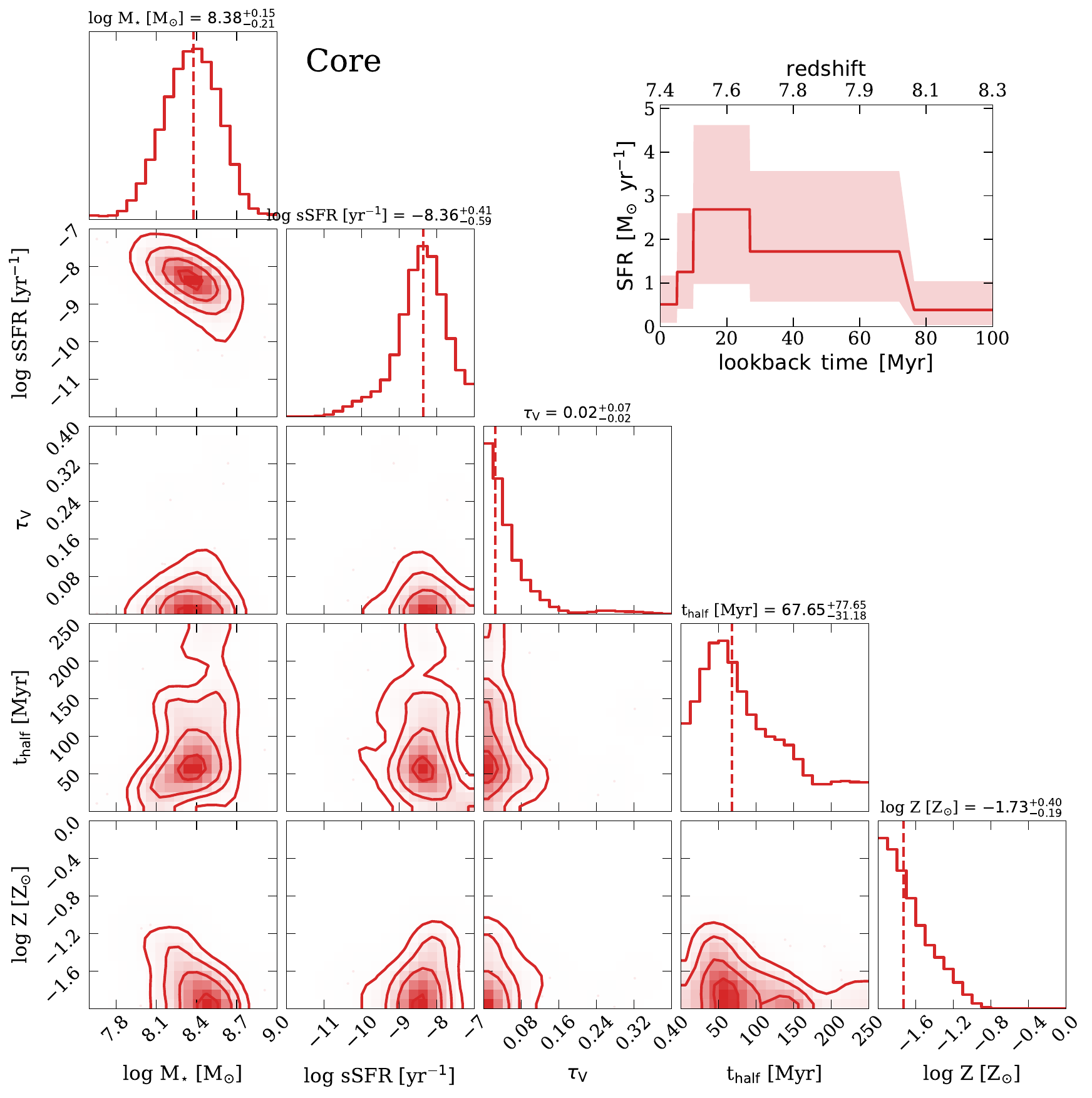}
    \caption{\textbf{Corner figure and star-formation history for the core component.} Left: Corner plot showing stellar mass ($ M_*$), specific star-formation rate (sSFR), optical depth ($\tau_\nu$), half-time ($t_{\rm half}$) and stellar metallicity (Z) for the core component as obtained by SED fitting. Right: the SFH for the core component. The data is presented as the median of the distribution with the errors corresponding to the 16th and 84th percentiles. }
    \label{fig:core_corner}
\end{figure}

\begin{figure}
    \centering
    \includegraphics[width=\columnwidth]{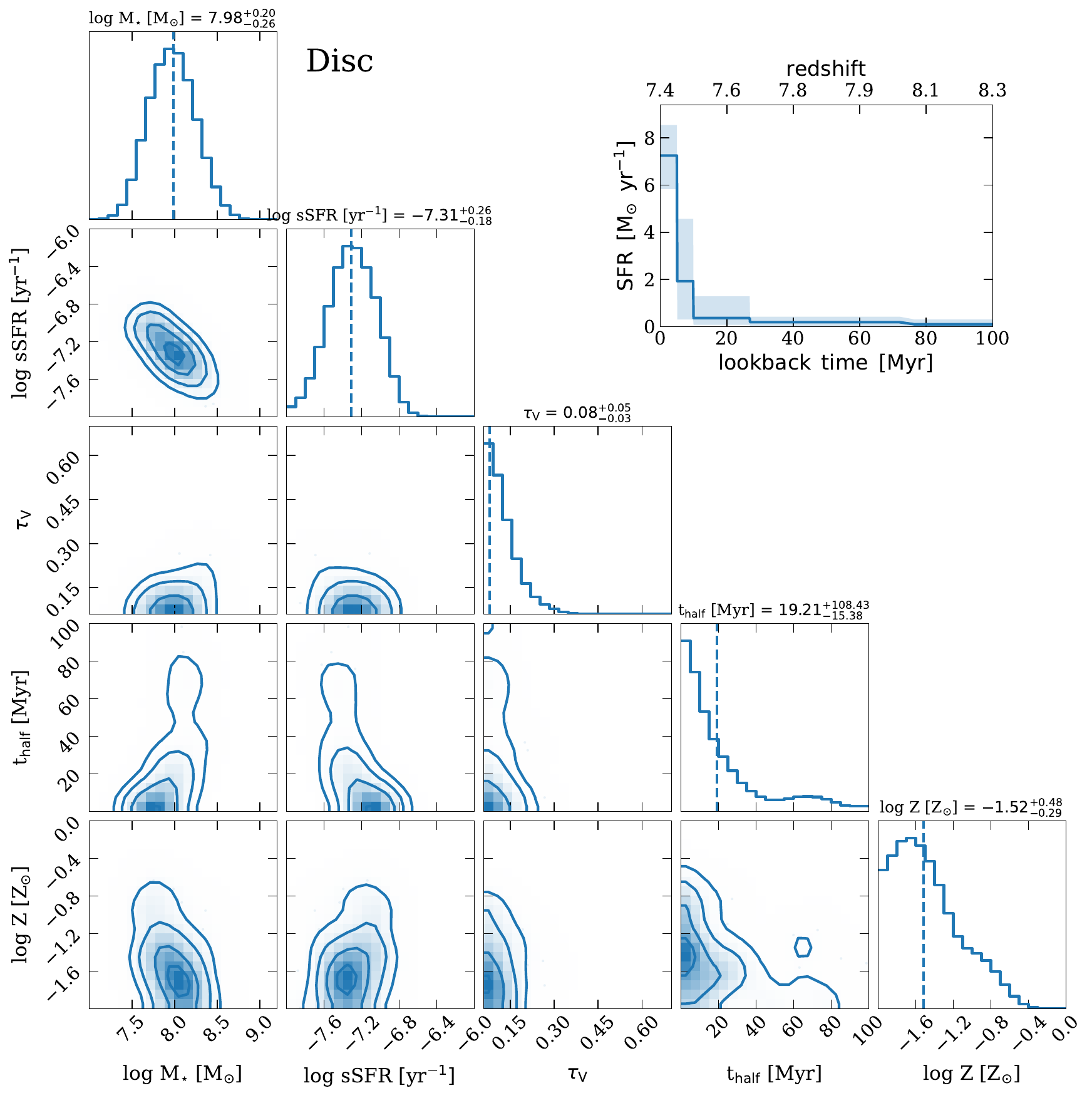}
    \caption{\textbf{Corner figure and star-formation history for the disc component.} Left: Corner plot showing stellar mass ($ M_*$), specific star-formation rate (sSFR), optical depth ($\tau_\nu$), half-time ($t_{\rm half}$) and stellar metallicity (Z) for the disc component as obtained by SED fitting. Right: the SFH for the disc component. The data is presented as the median of the distribution with the errors corresponding to the 16th and 84th percentiles.}
    \label{fig:disc_corner}
\end{figure}

\begin{figure}
    \centering
    \includegraphics[width=\columnwidth]{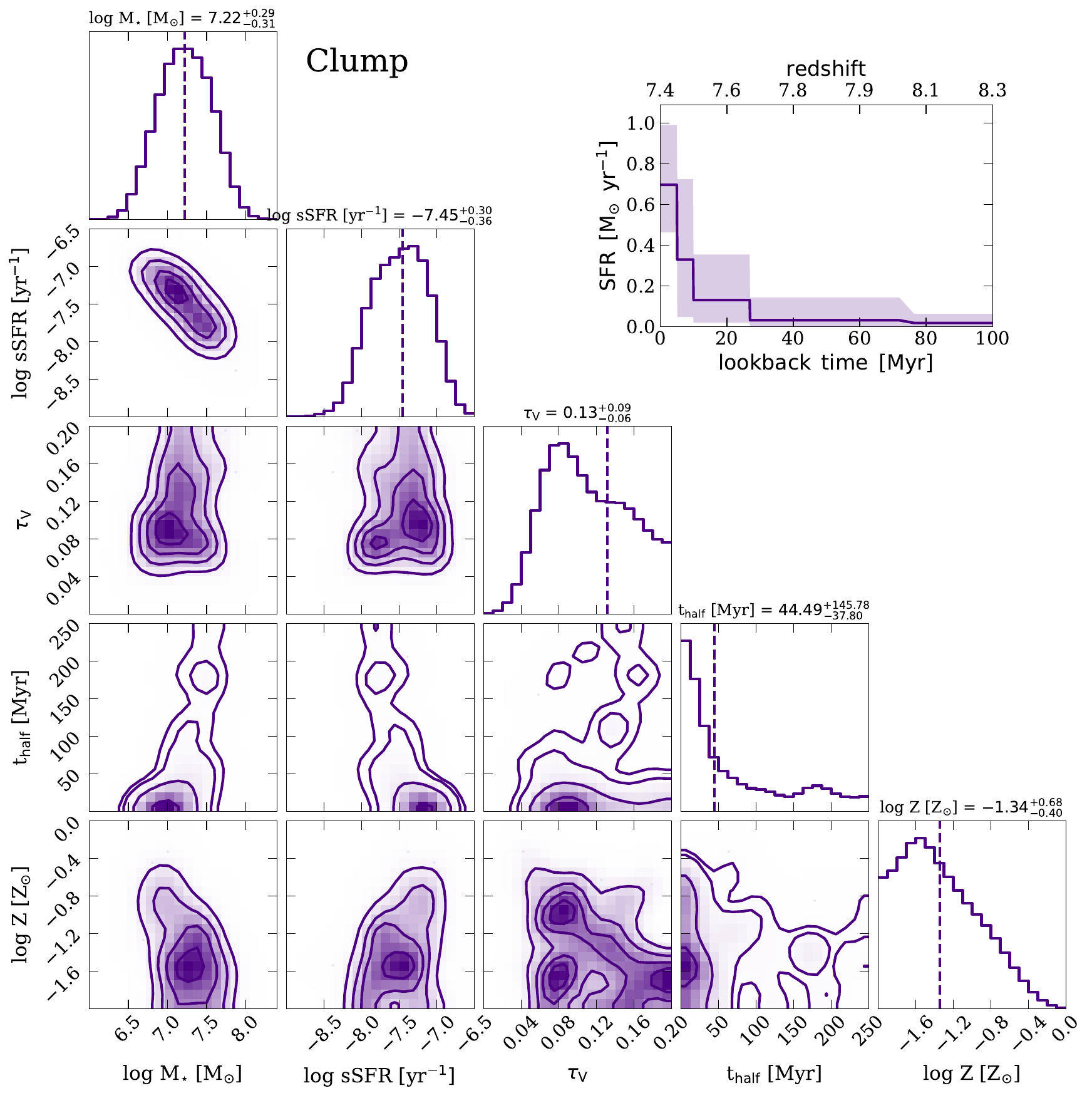}
    \caption{\textbf{Corner figure and star-formation history for the clump component.} Left: Corner plot showing stellar mass ($ M_*$), specific star-formation rate (sSFR), optical depth ($\tau_\nu$), half-time ($t_{\rm half}$) and stellar metallicity (Z) for the clump component as obtained by SED fitting. Right: the SFH for the clump component. The data is presented as the median of the distribution with the errors corresponding to the 16th and 84th percentiles.}
    \label{fig:clump_corner}
\end{figure}

Figures \ref{fig:core_corner}, \ref{fig:disc_corner}, and \ref{fig:clump_corner} show the corner plots for each component for the stellar mass, sSFR, optical depth of the diffuse component, stellar age (lookback time at which 50\% of the stellar mass was formed), and stellar metallicity. The SFHs are shown on the top right. Despite not fully breaking the dust-age-metallicity degeneracy, we are able to constrain the stellar mass and overall SFH well. 

In order to explore the dependency of our results on the SFH prior, we also tried the bursty-continuity prior (\cite{Tacchella2022b}), which enables the SFH to change more rapidly, enabling a more variable (i.e. bursty) star-formation history. In the bursty continuity prior case we obtain stellar masses of  $\rm log(M_*/M_\odot)=8.49^{+0.22}_{-0.35}$, $\rm log(M_*/M_\odot)=7.75^{+0.18}_{-0.09}$ and $\rm log(M_*/M_\odot)=6.89^{+0.37}_{-0.14}$ for the core, disc and clump components, respectively, compared to $\rm log(M_*/M_\odot)=8.48^{+0.16}_{-0.21}$, $\rm log(M_*/M_\odot)=8.05^{+0.23}_{-0.29}$ and $\rm log(M_*/M_\odot)=7.29^{+0.34}_{-0.36}$ in the standard continuity prior case. Therefore, the stellar masses obtained in both cases are consistent within the errors, suggesting that the stellar masses obtained based on the standard continuity prior are not biased by failing to account for particularly bursty SFHs.

\subsection{SED fitting of the combined photometry}
\label{sec:multi_SED}

\begin{figure}
    \centering
    \includegraphics[width=0.6\columnwidth]{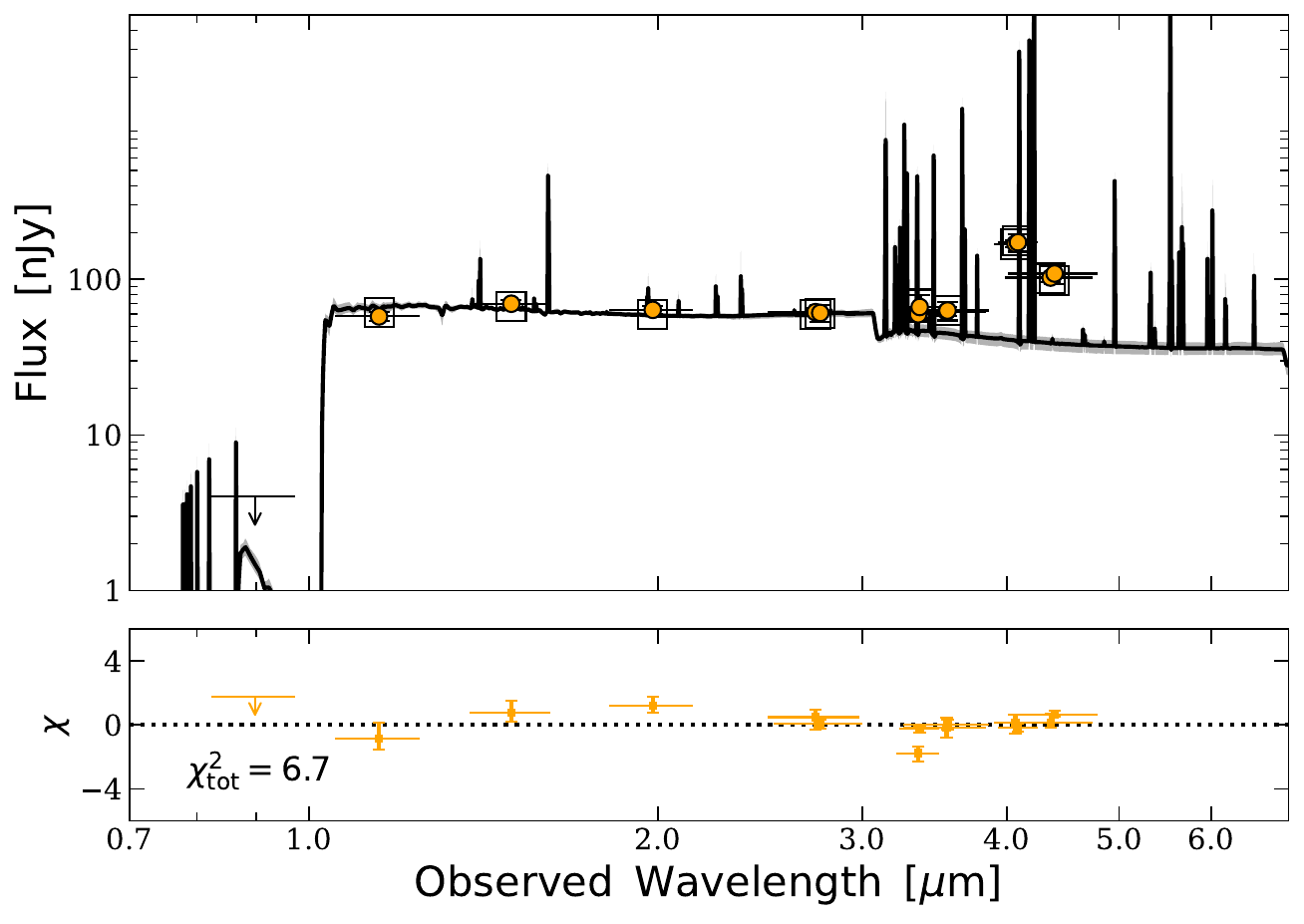}
    \includegraphics[width=\columnwidth]{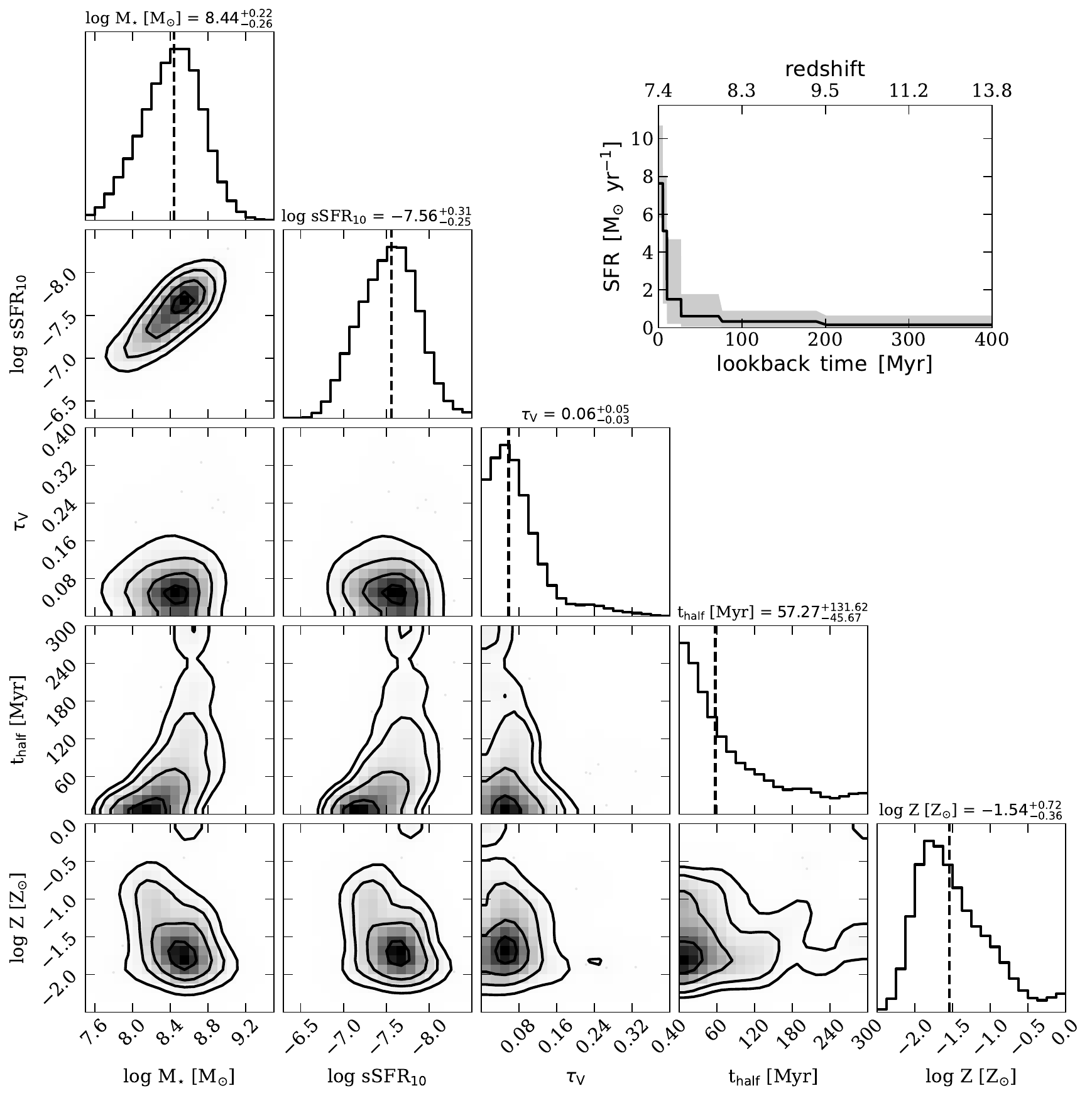}
    \caption{\textbf{SED fit, corner figure and SFH for the combined photometry.} Upper panel: Spectral energy distribution (SED) fits for the single-component ForcePho fit. The yellow points show the photometry inferred from the ForcePho modelling, while the 2$\sigma$ upper limits are indicated as downward pointing arrows. The errors correspond to the 1$\sigma$ uncertainties from the photometric pipeline. The open squares mark the photometry from the best-fit SED model. The solid lines and the shaded regions show the median and the 16th-84th percentile of the SED posterior from the Prospector modelling, respectively. Lower panel: corner plot showing stellar mass ($ M_*$), specific star-formation rate (sSFR), optical depth ($\tau_V$), half-time ($t_{half}$) and stellar metallicity (Z) for the single component ForcePho fit as obtained by SED fitting, with the SFH inset. These two plots show that when fit as single component, the stellar mass and sSFR inferred, traces that of the combined galaxy.}
    \label{fig:one_comp_sed_fit}
\end{figure}

In order to assess how this galaxy relates to other galaxies, we need to infer the global stellar populations parameters from the combined photometry, i.e. treating the core, disc, and clump as a single SED. The combined SED can be seen in the top panel of Fig. \ref{fig:one_comp_sed_fit}, while the bottom panel shows the corner plot with the SFH inset. 

We obtain a stellar mass of $\rm log(M_*/M_\odot)=8.44^{+0.22}_{-0.26}$ and $\rm SFR_{10Myr}=6.3^{+1.6}_{-1.1}\,M_\odot\,yr^{-1}$ and a sSFR of $\rm log(sSFR/yr)=-7.56^{+0.31}_{-0.25} \,yr^{-1}$. 
For comparison, the combined stellar masses of the individual components is $\rm log(M_*/M_\odot)=8.65^{+0.25}_{-0.30}$, while the SFR amounts to $\rm SFR_{\rm 10Myr}=11.5^{+5.7}_{-4.3}\,M_\odot\,yr^{-1}$. 
This means that the results are consistent within the uncertainties quoted. We show the results of fitting this combined photometry in Fig. \ref{fig:sfms} as the orange marker. We see that it is consistent with the black marker (the results of adding the stellar masses and SFRs of the core and disc components). 
In summary, we find that this galaxy increases its SFH, as expected for galaxies at this epoch.

\subsection{The clump component}

\begin{figure}
    \centering
    \includegraphics[width=0.45\columnwidth]{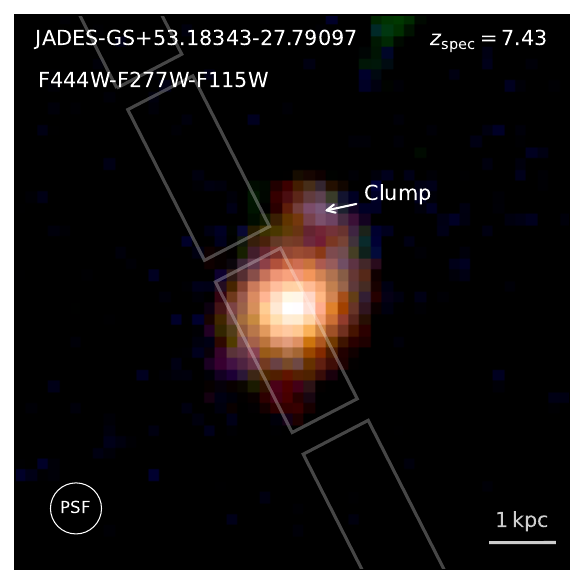}
    \includegraphics[width=0.54\columnwidth]{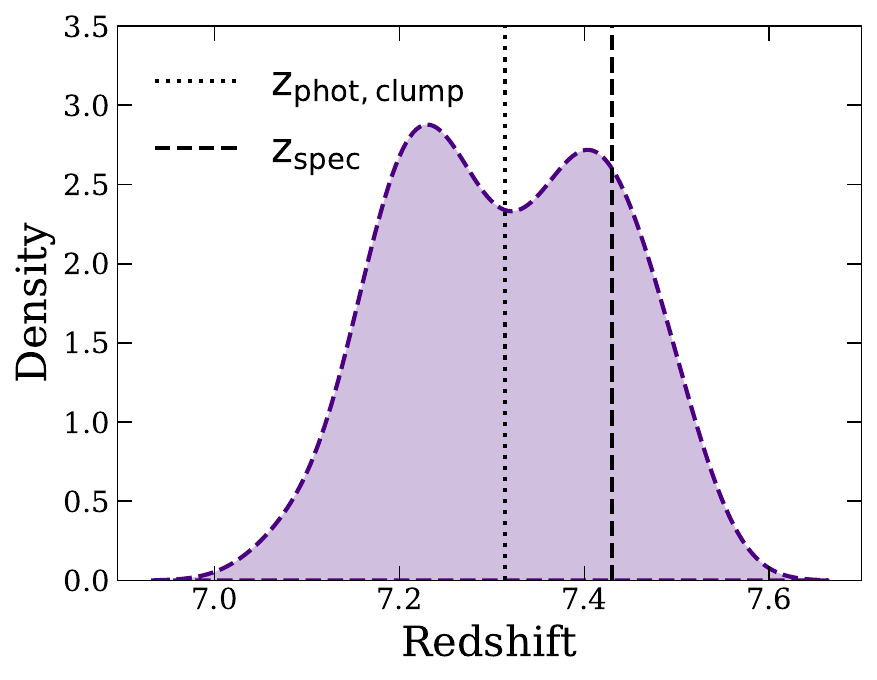}
    \caption{\textbf{RGB image highlighting the clump and 1D posterior distribution for the photometric redshift of the clump component.} Left: RGB image of \jnamespace with the clump position highlighted. Right: the marginalised posterior distribution for the photometric redshift of the clump component. We see it is double peaked but consistent with the spectroscopic redshift of the core and disc galaxy.}
    \label{fig:clump_rgb_posterior}
\end{figure}

In this section, we explore the clump component seen in the imaging data and modelled as a point source. The left panel of Fig. \ref{fig:clump_rgb_posterior} shows the RGB image of the galaxy with the position of the clump highlighted. The clump itself does not fall into the NIRSpec slit, so it is natural to wonder if it is actually at the same redshift as the core-disc components. Based on the RGB image, it appears to have similar colours, suggesting that this is likely. To quantify this further, we determine a photometric redshift by running Prospector on the ForcePho photometry with redshift left as a free parameter with a top-hat prior varying from 0.1-13. Figure \ref{fig:clump_rgb_posterior} shows the resulting density distribution for the redshift of the clump component. It appears to be double-peaked, but is consistent with the core and disc component's spectroscopic redshift. This means that it is realistic to consider that the clump is either part of the galaxy (i.e. a violent disc instability) or a merging galaxy at the same redshift.

\subsection{Emission line properties}
\label{sec:em_lines}

\begin{figure}
    \centering
    \includegraphics[width=0.6\columnwidth]{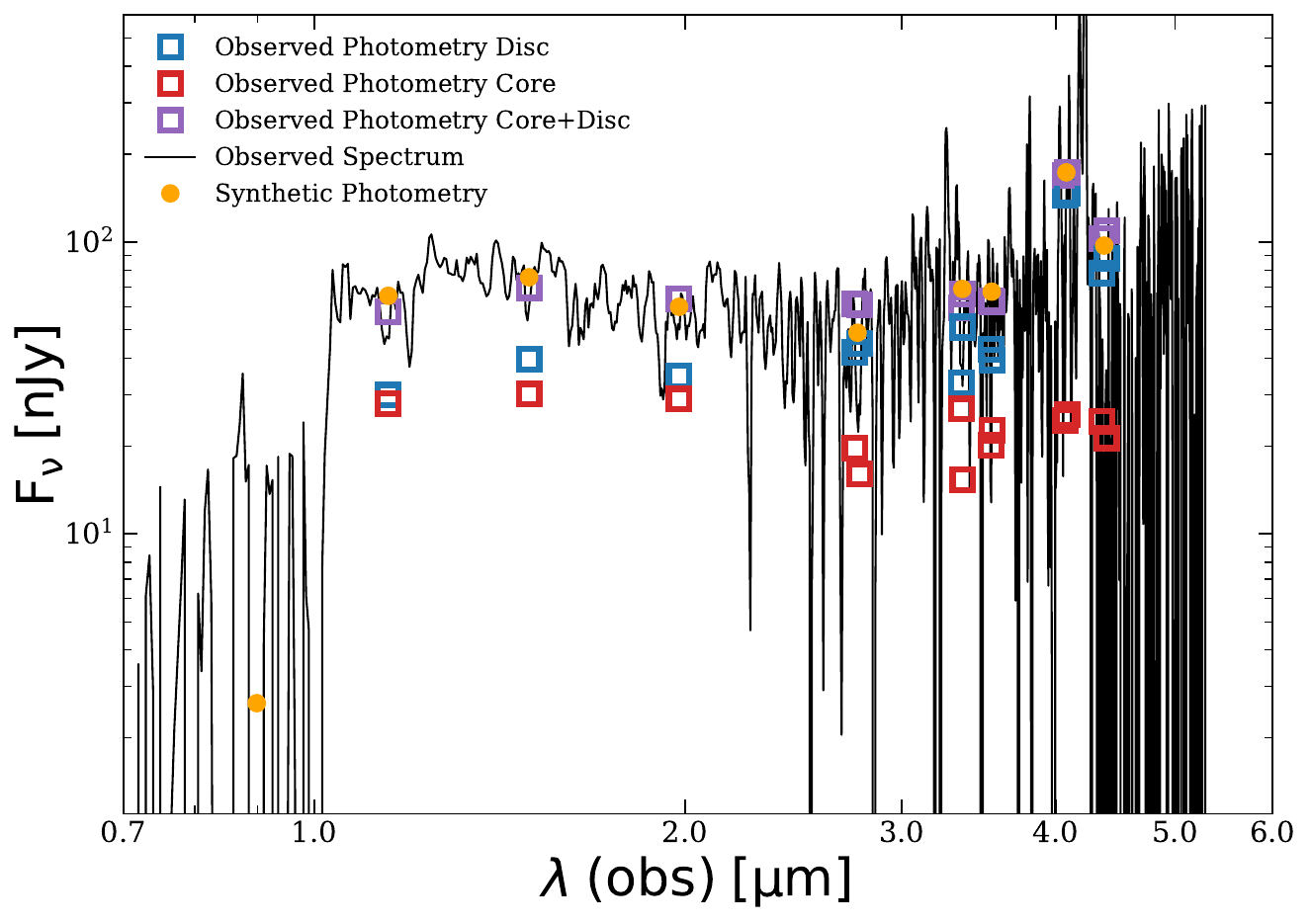}
    \caption{\textbf{The NIRSpec spectrum compared to the ForcePho photometry.} The 1D NIRSpec spectrum (black) and synthetically derived photometry from the spectrum (orange points). The observed ForcePho photometry for the disc (blue), core (red) and combined core+disc galaxy (lilac) is overplotted. We find the synthetic photometry matches the photometry of the combined core + disc galaxy, highlighting that the combined medium- and broad-band photometry traces both the stellar continuum and nebular emission line.}
    \label{fig:spec_fnu}
\end{figure}

The NIRSpec R100 spectrum contains crucial information. However, as outlined above, we only include the inferred redshift as a constraint when performing SED modelling of the morphologically distinct components. The main reason for this is that the spectrum (see Fig.~\ref{fig:rgb}) covers only parts of the galaxy and includes both core and disc components, but the degree of which is unknown. In principle, one could forward model the full ForcePho model through the slit, then perform a simultaneous fit of both SED components. Unfortunately, this is not yet possible with the Prospector framework.

Nevertheless, it is important to compare whether the photometry and spectroscopy are consistent with each other and to compare the SED derived quantities with the ones obtained only from the spectrum. Figure \ref{fig:spec_fnu} shows the NIRSpec spectrum (black) with synthetic photometry (orange) produced by integrating the flux from the spectrum within each filter. The observed photometry for the core (red) component, the disc (blue), and the combined core+disc galaxy (lilac) is overplotted. The key comparison here is with the combined photometry (in lilac). We see that the synthetic photometry follows the same trends as the combined \texttt{ForcePho} photometry with strong jumps flux in F410M and F444W corresponding to the strong line emission in the galaxy. We use the measured flux from the $\rm [OIII]\lambda5007,4963$ doublet and the H$\beta$ emission line from the spectrum to calculate their contribution to the F410M and F444W filters. We find that they account for 46\% of the flux in the F410M band and 23\% of the flux in F444W, hence both of these bands are being boosted by the emission line flux. We also see that the overall shape of the synthetic photometry and the actual spectrum mirror the actual photometry with a median $\chi^2$ value of 2.5. If we compare the values of the emission lines from the spectrum to the photometry, we find that they are in good agreement (within $1\sigma$) with values of $\rm [OIII]\lambda5007,4963=7.52_{-0.12}^{+0.10} [10^{-18} erg\;cm^{-2}\;s]$ and $\rm H\beta=0.91_{-0.08}^{+0.11} [10^{-18} erg\;cm^{-2}\;s]$ from the observed NIRSpec spectrum and $\rm [OIII]\lambda5007,4963=8.45_{-3.33}^{+0.6.86} [10^{-18} erg\;cm^{-2}\;s]$ and $\rm H\beta=1.89_{-0.74}^{+0.1.87} [10^{-18} erg\;cm^{-2}\;s]$ for the combined core+disc galaxy from the best-fit photometry.

We infer several important quantities from the emission lines obtained from the NIRSpec R100 prism spectra (see the bottom panel of Fig. \ref{fig:rgb}). 
First, we do the fitting and measure the fluxes (Section \ref{sec:spec}), then we correct the fitted emission lines for extinction using the ratio between the H$\gamma$ and H$\beta$ Balmer lines (as in \cite{Curti2023a}). The intrinsic value of the ratio (assuming Case B recombination, electron temperature $T_e=1.5\times 10^4$ K and an electron density of $N_e=300~\mathrm{cm}^{-3}$) is 0.468. We measure H$\gamma$/H$\beta$=0.409 meaning we are seeing the effects of dust. We obtain a dust extinction of A$_{\rm V, gas}=0.8_{-0.8}^{+1.2}$ mag in the V band. 
We also note that this dust extinction is also consistent with that obtained from our SED fitting (of only the photometry), where we obtain $A_{\rm V}=0.31^{+0.20}_{-0.13}$ mag for the disc.

We calculate the gas-phase metallicity of \jnamespace using the strong line method (e.g., \cite{Curti2017,Bian2018,Curti2020, Baker2023b, Baker2023c}) which uses the ratios of strong emission lines, in this case [OIII]$\lambda$5007, [OII]$\lambda$3727, H$\beta$, and [NeIII]$\lambda$3969. We use the strong-line metallicity diagnostics from \cite{Bian2018}.

We obtain a gas-phase metallicity of 12+log(O/H)=$7.86^{+0.09}_{-0.09}$, broadly consistent with that of other galaxies at these redshifts \cite{Sanders2024}. This is equivalent to a value of log(Z$_{\rm gas}$/Z$_\odot$) = -0.83, which is larger than the stellar metallicities inferred for the core and disc from Prospector, but broadly consistent with the average of the gas-phase metallicity inferred for the two with log(Z$_{\rm gas}$/Z$_\odot$)=$-1.46^{+0.72}_{-0.37}$ for the disc and log(Z$_{\rm gas}$/Z$_\odot$)=$-0.955^{+0.54}_{-0.26}$ for the core. 

We can calculate an estimate of the SFR from the dust-corrected H$\beta$ emission line by assuming a Balmer decrement flux ratio of $\rm F_{H\alpha}/F_{H\beta}=2.86$, corresponding to case B recombination at a temperature of $\rm T\sim 10^4$K (as in \cite{Baker2022, Curti2023b}). This enables us to estimate H$\alpha$. We then convert this H$\alpha$ flux into a H$\alpha$ luminosity, which we convert into a SFR using the conversion detailed in \cite{Shapley2023}. This gives us a SFR of $\rm SFR=8.5^{+34.7}_{-7.7}\;M_\odot\,yr^{-1}$. This is consistent with the combined SFR of the core and disc obtained via SED fitting with Prospector where $\rm SFR_{\rm 10Myr}=11.5^{+5.7}_{-4.3}\,M_\odot\,yr^{-1}$.

\subsection{Star formation vs AGN}
\label{sec:SF_AGN}

\begin{figure}
    \centering
    \includegraphics[width=0.45\columnwidth]{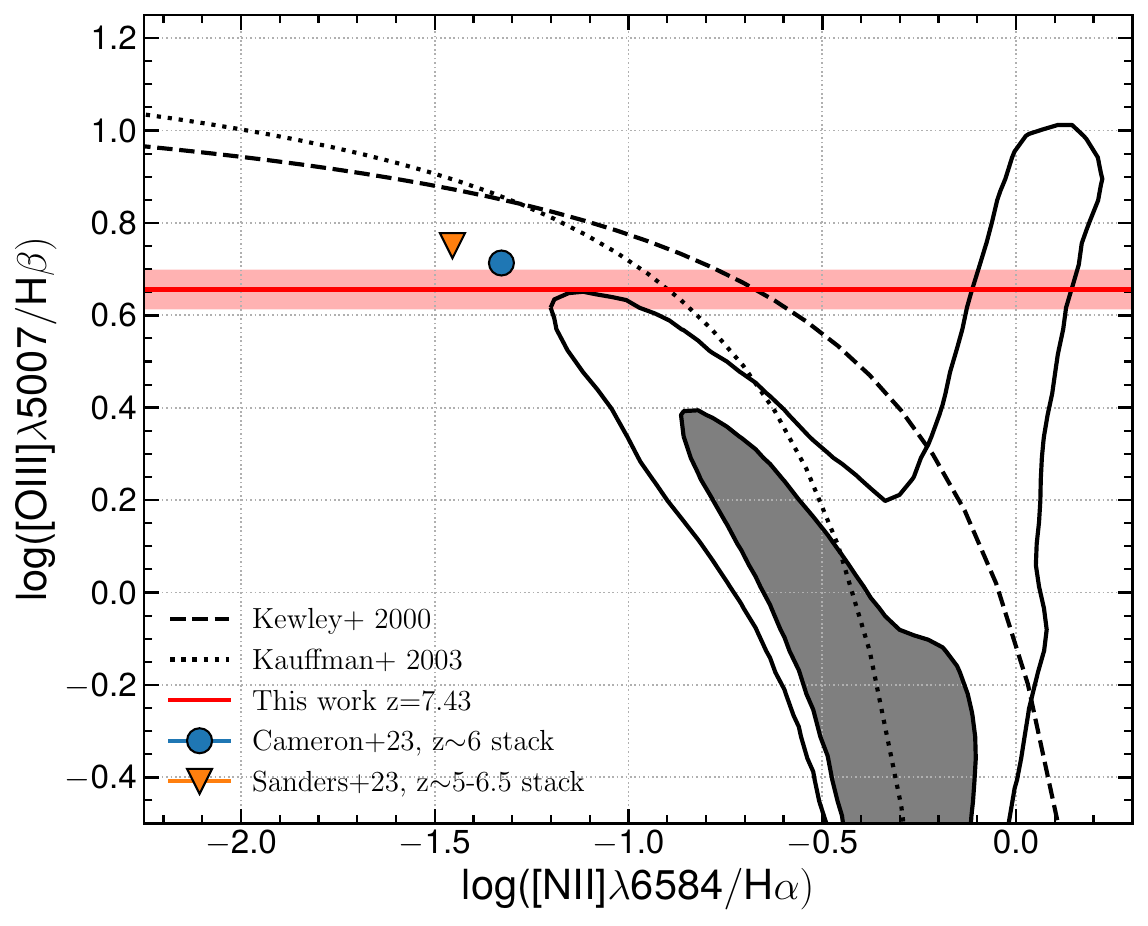}
    \includegraphics[width=0.49\columnwidth]{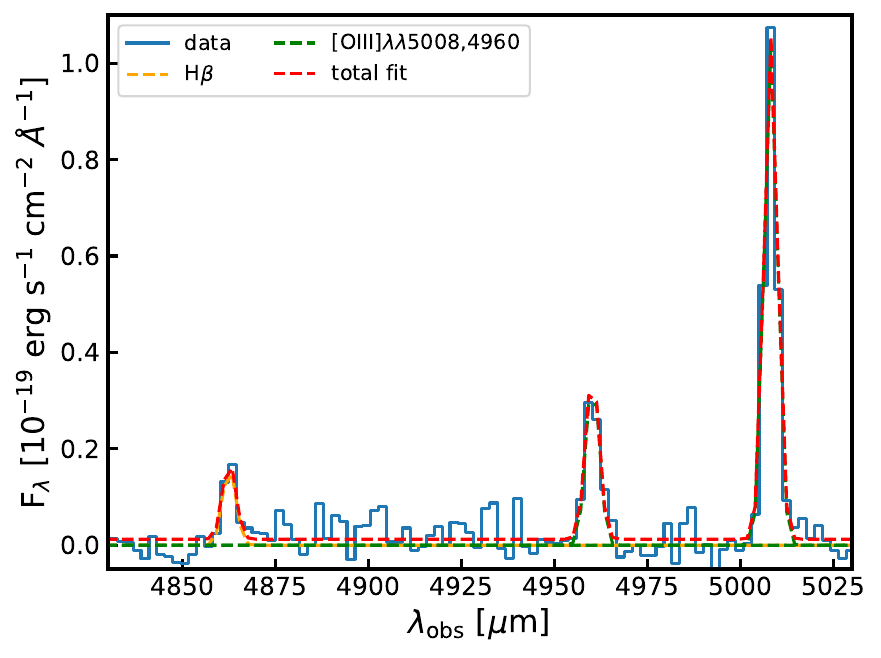}
    
    \caption{\textbf{Emission line diagnostic diagram and examination of fits to the R1000 spectrum.} Left panel: A classical BPT \cite{Baldwin1981} diagram showing the ratio of the emission lines [OIII]$\lambda$5007 to H$\beta$ versus $[NII]\lambda6584$ / H$\alpha$. At this redshift NIRSpec can no longer detect $[NII]\lambda6584$ and H$\alpha$ so we plot a red horizontal line for the value of [OIII]$\lambda$5007/H$\beta$, the colour fill corresponds to the $1\sigma$ error propagated through from the pipeline. We see that the galaxy is still consistent with a star-forming galaxy and shows similar [OIII]$\lambda$5007/H$\beta$ ratios to other star-forming galaxies at these ratios \cite{Cameron2023}. 
    Right panel: fits to the H$\beta$ emission line and the [OIII] doublet. We see that both can be with standard Gaussians and that there is no need for an underlying broad component.}
    \label{fig:bpt}
\end{figure}

Is the central component a stellar core or an AGN? We know that the central component of this galaxy is compact, with a deconvolved half-light radius of about 80 pc and a S\'ersic index of 2.0. We use dust corrected emission line diagnostics from the NIRSpec spectrum to investigate a possible AGN contribution. We use the ratio [OIII]$\lambda$5007/H$\beta$ from the classical BPT \cite{Baldwin1981} diagram and find that this gives us a value of $\sim$4.5 (see Fig. \ref{fig:bpt}), which, while large for the local Universe, is consistent with star-formation at high redshifts \cite{Cameron2023, Shapley2023b}. Fig. \ref{fig:bpt} left panel shows the BPT diagram of [OIII]$\lambda$5007/H$\beta$ against [NII]$\lambda$6584/H$\alpha$. As \jnamespace has a redshift of 7.43, the [NII]$\lambda$6584 and H$\alpha$ emission lines are shifted out of the NIRSpec wavelength coverage, hence we plot the value of [OIII]$\lambda$5007/H$\beta$ as a straight red line. We also show the combined data stacks of \cite{Cameron2023} and \cite{Sanders2024}. We overplot contours from local SDSS galaxies with data from \cite{Baker2023b}.
We see that the line ratio for \jnamespace appears consistent with those of the stacks for galaxies at similar redshifts, but this itself cannot rule out an AGN contribution (as shown in \cite{Scholtz2023}). Similarly, we do not find indications for strong high-ionisation emission lines such as He$\lambda$4686 that cannot be explained by stellar emission. Fig. \ref{fig:bpt} right panel shows line fits to the H$\beta$ and [OIII] doublet. The key takeaway from this is that we see no broadening of the H$\beta$ line compared to the [OIII]. This means \jnamespace shows no evidence of being a type-1 AGN. 

Overall, this shows that \jnamespace appears to be consistent with pure star formation, but we note that it is difficult to fully rule out an AGN.

\section*{Data availability}
The JADES data is publicly available at \url{https://jades-survey.github.io/scientists/data.html} or through the Mikulski Archive for Space Telescopes (MAST) \url{https://archive.stsci.edu/hlsp/jades}.
Additional data derived from the raw products is available from the corresponding author upon reasonable request. 

\section*{Code availability}
AstroPy \cite{Astropy_2022ApJ...935..167A}, Prospector \cite{Prospector2021}, Dynesty \cite{Speagle2020}, FSPS \cite{Conroy2009, Conroy2010}, Galsim \cite{Galsim_2015}, WebbPSF and Photutils \cite{Bradley_photutils_2022zndo...7419741B}, are all publicly available, while ForcePho (Johnson et al. in prep) is publicly available via GitHub at \url{https://github.com/bd-j/forcepho}.

\section*{Acknowledgements}

ST acknowledges support by the Royal Society Research Grant G125142.
WB, TJL, FDE, RM, JW, LS and JS acknowledge support by the Science and Technology Facilities Council (STFC) and by the ERC through Advanced Grant 695671 “QUENCH”. 
RM also acknowledges funding from a research professorship from the Royal Society. 
JW further acknowledges support from the Fondation MERAC. This study made use of the Prospero high performance computing facility at Liverpool John Moores University. 
BDJ, EE, MR, BER and CNAW acknowledge support from the JWST/NIRCam Science Team contract to the
University of Arizona, NAS5-02015. 
ECL acknowledges support of an STFC Webb Fellowship (ST/W001438/1). 
SC acknowledges support by European Union’s HE ERC Starting Grant No. 101040227 - WINGS. 
AJB, JC acknowledge funding from the "FirstGalaxies" Advanced Grant from the European Research Council (ERC) under the European Union’s Horizon 2020 research and innovation programme (Grant agreement No. 789056).  
SA acknowledges support from the research project PID2021-127718NB-I00 of the Spanish Ministry of Science and Innovation/State Agency of Research (MICIN/AEI). 
H{\"U} gratefully acknowledges support by the Isaac Newton Trust and by the Kavli Foundation through a Newton-Kavli Junior Fellowship. 
DJE is supported as a Simons Investigator and by JWST/NIRCam contract to the University of Arizona, NAS5-02015. 
D.P. acknowledges support by the Huo Family Foundation through a P.C. Ho PhD Studentship.
A.L.D. thanks the University of Cambridge Harding Distinguished Postgraduate Scholars Programme and Technology Facilities Council (STFC) Center for Doctoral Training (CDT) in Data intensive science at the University of Cambridge (STFC grant number 2742605) for a PhD studentship.
The reserach of CCW is supported by NOIRLab, which is managed by the Association of Universities for Research in Astronomy (AURA) under a cooperative agreement with the National Science Foundation.
The authors acknowledge use of the lux supercomputer at UC Santa Cruz, funded by NSF MRI grant AST 1828315.
Funding for this research was provided by the Johns Hopkins University, Institute for Data Intensive Engineering and Science (IDIES). 
This research is supported in part by the Australian Research Council Centre of Excellence for All Sky Astrophysics in 3 Dimensions (ASTRO 3D), through project number CE170100013.

\section*{Author Contributions Statement}

WMB and ST led the writing of the paper. 
WMB performed the analysis (identification of the target, ForcePho fitting, SED modelling, and figure making) under the supervision of ST and BDJ.
All authors have contributed to the interpretation of the results.
DJE, BDJ, BR, ST, DP, RH, ZJ, and CNAW contributed to the NIRCam imaging reduction.
EN, KS, DJE, BDJ, BR, and ST contributed to the analysis and interpretation of the NIRCam imaging data. 
FDE contributed to the development of tools for the spectroscopic data analysis. SC, SA, MC, and JW contributed to the reduction of NIRSpec data and the development of the NIRSpec pipeline. AB contributed to the design and optimisation of the MSA configurations. FDE, MC, ADG, RM and JS helped with the interpretation of the NIRSpec data.

\section*{Competing Interests Statement}
The authors declare that they have no competing interests.

\bigskip



\bibliographystyle{sn-nature}
\bibliography{refs.bib}

\end{document}